%% file: ms.tex
\documentstyle[11pt,aaspp4,psfig]{article}      

\include{definition}

\begin{document}

\title{A New Study of s-Process Nucleosynthesis in Massive Stars}
\centerline{ I. The Core-Helium Burning Phase}
\author{L.-S. The,$^1$   M.F. El Eid,$^2$   and   B.S. Meyer$^1$}
\affil{$^1$Department of Physics \& Astronomy, Clemson University,
       Clemson, SC 29634-1911 }
\affil{$^2$Department of Physics, American University of Beirut (AUB),      
       Beirut, Lebanon}

\begin{abstract}

We present a comprehensive study of s-process nucleosynthesis in
15, 20, 25, and 30 $\msun$ stellar models having solar-like
initial composition. The stars are evolved up to ignition of central
neon with a 659 species network coupled to the stellar
models. In this way, the initial composition from one burning phase to
another is consistently determined, especially with respect to neutron
capture reactions.  The aim of our calculations is to gain
a full account of the s-process yield from massive stars. 
In the present work, we focus primarily on the s-process during central helium
burning and illuminate some major uncertainties affecting the calculations. 
We briefly show how advanced burning can significantly affect the products
of the core helium burning s-process and, in particular, can greatly deplete
$^{80}$Kr that was
strongly overproduced in the earlier core helium burning phase; however,
we leave a complete analysis of the s-process during the
advanced evolutionary phases (especially in shell carbon burning)
to a subsequent paper.  Our results can help to constrain the yield of the
s-process material from massive stars during their pre-supernova
evolution.
\end{abstract}

\keywords{nuclear reactions, nucleosynthesis, abundances 
          --stars: evolution --stars: interiors}

\newpage
\section{INTRODUCTION}

According to phenomenological analysis (e.g. \cite{C61},
\cite{K89}),
at least two distinct components are needed to fit the solar abundance
distribution of the s-process nuclei which is empirically described by the  
well-known $\sigma$N$_s$(A) curve.  There is the  weak s-process component
which is confined to the atomic mass range A$\sim$ 65--90 (from copper
to zirconium), and the main s-process component which is responsible for
the production of the heavier nuclei with A=90-- 209 (i.e. elements
up to bismuth).
These two components are associated with distinct stellar sites of the
s-process: the weak component is mainly synthesized in massive stars
(M$\ge 13\msun$), while the main component is produced during recurrent
helium-shell thermal pulses in AGB stars.

Our concern in this paper is a careful study of the weak s-process in
massive stars. In addition to its intrinsic interest in nucleosynthesis
theory, such a study is important for at least two other main reasons.
First, a well determined yield from the weak component helps to constrain
the contribution of the main component to the production of the nuclei in 
the mass range A=65--90.  This will help determine the degree to which s-process
isotopes in mainstream SiC grains should differ from solar ratios
(e.g. \cite{Gallino97}).
Secondly, the weak component depends strongly on 
the initial metallicity (e.g Prantzos, Nomoto, \& Hashimoto 1990;
\cite{B92}; \cite {R92}); thus, it may 
be used to study the role of massive stars in the early phase of the chemical
evolution of the Galaxy.

\cite{P68} was the first to suggest that the main neutron source triggering 
the s-process in massive stars is the $^{22}{\rm Ne}(\alpha,n)^{25}{\rm Mg}$
reaction. The $^{22}$Ne abundance is built up at the beginning of central
helium burning by the reaction chain
$^{14}{\rm N}(\alpha,\gamma)^{18}{\rm F}(e^+\nu_e)^{18}{\rm O}(\alpha,\gamma)
^{22}{\rm Ne}$, where the $^{14}$N is left over from CNO cycling during the
preceding hydrogen burning phase.  As the temperature rises in the helium
burning core, the reaction $^{22}{\rm Ne}(\alpha,n)^{25}$Mg releases neutrons
which may then capture onto $^{56}$Fe seed nuclei.

There are a number of key factors that influence the s-process
occurring in this site: 
\begin{enumerate}       
\item The $^{25}$Mg produced by $^{22}{\rm Ne}(\alpha,n)^{25}$Mg
serves (along with other nuclei such as $^{12}$C,
$^{20}$Ne, $^{16}$O) as a neutron poison, thereby reducing the number
of neutrons per $^{56}$Fe seed.

\item The reaction $^{22}{\rm Ne}(\alpha,n)^{25}{\rm Mg}$
becomes effective only during the late phase of central helium burning
when the core temperatures exceeds $\sim$2.5$\times 10^8$ K.
Since helium ignites at temperatures
less than $2 \times 10^8$ K, only a terminal
fraction of the core helium burning phase
actively contributes to the s-process in massive stars. 

\item The production of neutrons by the $^{22}$Ne
source is also limited by the efficiency of the reactions competing for the
alpha particles such as
the $^{12}{\rm C}(\alpha,\gamma)^{16}$O reaction, which has a rate still
not accurately determined to better than a factor of two,
and the $^{22}{\rm Ne}(\alpha,\gamma)^{26}{\rm Mg}$ reaction.

\item As our computations show (see \S 5), s-process nucleosynthesis
is  rather sensitive to the treatment of mixing at the edge of the convective
core during the final part of central helium burning 
(see discussion in sect. 5.1).  The yield depends also on the mass fraction
of the core that is ejected without significant additional nuclear modification.
We return to this theme at the end of this paper.
\end{enumerate}

Network calculations of the s-process require temperature, density, and
neutron abundance histories.  Due to computational limitations, past efforts
to model the s-process in massive stars have relied on the technique
known as ``post-processing'', specifically,
using temperature, density, and neutron abundance histories derived from
separately calculated stellar models. This may be a reasonable treatment  
of the s-process during central helium burning, but for advanced
burning stages (such as carbon-shell burning), rapid mixing and possible
feedback effects on the nuclear energy generation rate make the
post-processing approach questionable. In the calculations
by \cite{R91} and \cite{R93}, the contribution of
the carbon-burning shell to the s-process in a 25 $\msun$ star was calculated 
on the basis of the stellar models by \cite{N88}
by assuming a convective C-shell of constant temperature and density.
This is a simplified treatment in which a possible 
feedback effect on the stellar models and the important role of mixing
are not taken into account (as we show in \S \ref{sec:advanced}).

Given the complex behavior of the abundance pattern during the advanced
evolutionary phases,   
our goal is to avoid using post-processing to calculate      
the s-process in massive stars.
This is accomplished by coupling a full network of the s-process to the
stellar models (see \S 2 and \S 3). In this way, we are able to 
follow s-process products carefully
through the shell-helium burning and shell carbon burning phases in a
consistent way, and thereby to gain more insight into the s-process.
The results of our calculations are very rich in
detail; therefore, in order to keep the present paper manageable in scope, we
focus primarily on the core helium burning phase.  We do, however, present some
interesting results for carbon shell burning in \S \ref{sec:advanced}.

One of our primary goals in this work is to explore the effects of key
unresolved input uncertainties on the massive-star s-process yields.
Differences in s-process yields between massive star models largely result
from differences in the treatment of mixing within the stellar models
and from the uncertainty in
the $^{12}$C($\alpha$,$\gamma$)$^{16}$O rate (a factor of two) and
the neutron-capture cross sections in mass region between Fe and Zr
($\sim$10\% uncertainty).
We explore the effect of each of these uncertainties in
some detail.  To estimate the uncertainty and 
the affect of the $^{12}$C($\alpha$,$\gamma$)$^{16}$O rate 
on the s-process yields, we perform two calculations of a 
20 M$_{\sun}$ star, one using the rate given by Caughlan \& Fowler (1985)
(hereafter \cite{CFHZ85}) and the other
one using the rate given by Caughlan et al. (1988) (hereafter \cite{CF88}).
We discuss the differences in the stellar models by comparing our
results with those calculated by the FRANEC code (K\"appeler et al.
1994; hereafter \cite{K94}).  Comparison
of our results with those of Prantzos, Nomoto, \& Hashimoto (1990; hereafter
\cite{P90}) allows us to consider the effect
of the neutron capture cross section uncertainties. 

Additional uncertainties important for the weak s-process
arise from other key alpha particle
capture rates.  Recent progress in eliminating these uncertainties
has been made by \cite{K94} who measured
reaction rates for $^{18}$O($\alpha$,$\gamma$)$^{22}$Ne and
$^{22}$Ne($\alpha$,$\gamma$)$^{26}$Mg.
An uncertainty on the magnitude of the
$^{22}$Ne($\alpha$,n)$^{25}$Mg reaction
still remains, however.  In particular,
the possibility of a contribution to the rate from a
low-lying resonance at 633 keV in the compound nucleus
$^{26}$Mg has been shown to
increase the s-process production factors by an order of a magnitude above
the production factors from the rate given by \cite{CF88}.
In order to explore the effect of this uncertainty further,
we compare the s-process yields of
15, 20, 25, and 30 M$_{\sun}$ stars computed
with the $^{22}$Ne($\alpha$,n)$^{25}$Mg
rate given by \cite{CF88} and with the rate containing an additional
10\% contribution from the 633 keV resonance
(the WT case in \cite{K94}).

Despite the uncertainties discussed above,
the general notion of massive stars as the source the weak s-process
component is by now firmly in place.  A few difficulties remain, however.
Most significantly, a potential problem
has been recognized since the work of \cite{P90}.
In a comprehensive study, these authors calculated
s-process yields from massive stars during core helium burning as
a function of stellar mass and metallicity.
Feeding these as inputs into a simple chemical evolution model
and comparing the integrated abundance with solar abundance,
\cite{P90} concluded that massive stars during core helium burning
overproduce the nucleus $^{80}$Kr by a factor of $\simeq$2, thereby
presenting a serious difficulty for explaining the solar abundance of this
isotope.  One possible
solution considered by \cite{P90} is the destruction of $^{80}$Kr
during the supernova explosion.  This would probably only
destroy abundant $^{80}$Kr in the inner 2-3 solar masses.  This is
the matter that would for the most part remain in the stellar
remnant, however, and
thus the effect would
be unlikely to alter the ejected $^{80}$Kr abundance significantly.

Another solution of the $^{80}$Kr overproduction
problem has been suggested by Raiteri et al. (1991).
These authors argue that the solar s-only nuclei of 
the weak component originate not only from massive stars but
also in part from the main component (i.e. from thermal pulses in
the helium shell
of low-mass stars) and the p-process.
In particular, \cite{R91} find that massive stars contribute
significantly to the solar system's supply of
$^{80}$Kr through both the s- and p-processes.  Low-mass stars
do not.
On the other hand, they find that low-mass stars contribute abundantly
to the supply of other s-only nuclei in the $A\approx 60 - 90$ region.
In other words, they argue that the high production of $^{80}$Kr in massive
stars is needed to compensate for its relatively low production in low-mass
stars.  This is a plausible explanation but requires confirmation from
other calculations of the s-process in both low-mass and high-mass stars.

Based on the calculations presented here and in a 
forthcoming paper, we find an added wrinkle to the question of large
$^{80}$Kr overproduction factors
from massive star models.  We suggest that these overproduction
factors have been significantly overestimated because
calculations up to the present
have not yet correctly accounted for carbon-shell s-processing.  Our results
indicate that carbon shell burning can strongly deplete the large $^{80}$Kr
overabundances produced in core helium burning.  Interestingly, the amount
of depletion of $^{80}$Kr depends strongly on the amount of $^{22}$Ne left
over from core helium burning as well as the extent of and the timescale for
convective mixing in the carbon shell.  This identifies for $^{80}$Kr a
potentially powerful role as a diagnostic of advanced burning phases in
massive stars.  This is in addition to the important part it plays in
studies of galactic abundance evolution.

The outline of the present paper is the following.
In \S 2 we present an overview of
the stellar evolution code used to obtain the stellar models. In \S 3,
we briefly described the network we have used to calculate the s-process
abundances and to determine the energy generation rates in the stellar models.
In \S 4, \S 5 and \S 6 we summarize the results for the s-process, and in
\S 7 present our conclusions.

%
%
\section{STELLAR EVOLUTION CODE}

The stellar models presented in this work were computed with
a one-dimensional implicit hydrodynamical lagrangian code that solves
the stellar structure equations (see \cite{KW90}).
The original version of this code was described
in \cite{O83}. The present version of this code has the following general
features: the discretizied stellar structure  equations are iterated for a 
given time step in terms of five variables (radius r, velocity u,
temperature T, density $\rho$, and luminosity L). This is done for the whole
stellar model from the center to the photosphere with appropriate central and
surface boundary conditions. In particular, the surface boundary conditions are
adjusted to a gray atmosphere in a manner similar to that
described by \cite{P77}. The evolutionary time step is chosen by imposing 
upper limits on the variations of the physical variables in the stellar
model. The physical assumptions and parameterizations
used to construct the stellar models are 
summarized as follows:    
\begin{enumerate}
\item Rosseland  mean opacities from Livermore (\cite{IR96})
have been used for the hydrogen-rich stellar layers and for the
hydrogen-free, but C/O enhanced layers. The OPAL opacities are not
available for low temperatures (below about 6000 K). For such
temperatures, we used the Rosseland mean opacities calculated by \cite{AF94} 
which include the effects of molecules on the opacities.
Conductive opacities have also been included according to the analytical
approximation by \cite{Iben75}.

\item Partial ionizations of H, He, C, N, O, Ne, Mg and Si and their
effect on the equation of state are included by solving the 
Saha equation taking into account all relevant excited atomic states 
(\cite{EH90}). In fully ionized stellar layers, the 
equation of state includes a full relativistic description of the
degenerate electron and positron gas (\cite {CG68}). The effect of the
Coulomb pressure on the equation of state is approximately treated
according to the one-dimensional plasma calculations by
\cite{H73}.

\item The extensions of the convective regions in the stellar models
are determined according to the Schwarzschild criterion for convective
instability, $\nabla_{\rm rad} \geq \nabla_{\rm ad}$. We do not
include overshooting (e.g. \cite{S92}; \cite{B93}), 
or semiconvection (e.g. \cite{E95}, \cite{LM95}).
There is no consistent theory yet that describes these complex
effects, and their inclusion would complicate the discussion of the 
s-process, and prevent the comparison with the results of previous
calculations. 
Convective energy transport is treated according to the 
mixing length theory (see \cite{KW90}) with a mixing length parameter
l=2.0 H$_{\rm P}$, where ${\rm H_P}$ is the local pressure scale height.

\item Incrementing the abundances of nuclear species due to nuclear
reactions is followed by integrating a detailed network of nuclear
reactions as described in \S 3. The change of the abundances
due to convective mixing is performed by solving implicitly a
diffusion equation (see \cite{L85}; El Eid 1995) of the
following form:
\begin{equation}
{\partial X_i \over \partial t} = {\partial \over \partial M_r}
\left [ (4\pi r^2\rho)^2 D {\partial X_i \over \partial M_r} \right ],
\label{eq:dxidt}
\end{equation}
where $r$ is the radius, $\rho$ is the density, M$_{\rm r}$ is the mass
coordinate. The diffusion coefficient D in Eq.
(\protect\ref{eq:dxidt}) is taken as
${\rm D =v_c l}$, where ${\rm v_c}$ is the convective
velocity obtained from the mixing length theory, and ${\rm l}$ is the
mixing length parameter. It must be emphasized that mixing according to
Eq. (\ref{eq:dxidt}) does not always resemble instantaneous mixing. It does
so only if the nuclear time scale $\tau_{\rm nuc}$ is larger than
the diffusion time scale $\tau_{\rm diff}\sim l^2/D$. Such
conditions are encountered during the early evolution phases:
hydrogen, helium and carbon burning phases. Beyond the
central carbon phase, however, $\tau_{\rm nuc}$ may become comparable or
smaller than $\tau_{\rm diff}$, especially in the case of the light
particles (neutrons, protons and $\alpha$-particles) which are
liberated by nuclear reactions.  Neutrons set free are captured before any
mixing can occur.

The stellar models during these advanced phases were constructed
in the following way: for a given evolution time step $\Delta t$,
a stellar model is first constructed. With this time step, the 
network of nuclear reactions is integrated, and the resulting
abundances are mixed according to Eq. (\ref{eq:dxidt}). Subsequently, the
smallest diffusion time scale $\tau_{\rm diff}$ in the stellar
model is determined. If $\tau_{\rm diff}$ is found smaller than
$\Delta t$, the network is integrated again but with a fraction of
$\tau_{\rm diff}$ (typically $1 - 10\%$)
in order to correct the abundance profiles for
all nuclear species whose lifetimes are smaller than $\tau_{\rm
diff}$.  Through a number of tests,
we have convinced ourselves that this approximate
method of treating time-dependent mixing leads to the correct
energy generation rates in the stellar models.

\item In the present models,we have included mass loss by stellar
wind according to the semi-empirical relation by De Jager et al. (1988).
Mass loss for the stars in the mass range considered
here has only a minor effect on the central conditions affecting the 
s-process nucleosynthesis during core helium burning. 

Finally we mention that the neutrino energy losses due to pair, photo,
plasma, bremsstrahlung and recombination processes are taken into account
according to the recently up-dated analytical approximation by
Itoh et al. (1996).
\end{enumerate}

\section{THE REACTION NETWORK}

We have incorporated a large nuclear network into the stellar evolution
code in order to follow the s-process nucleosynthesis occurring during
advanced stages of core and shell burning.  The nuclear species included
in our network are those listed in table \ref{tab:tabnet}.  
Up to zinc, the network includes isotopes of each element ranging from 
an isotope near the proton-drip line to an isotope two mass units greater 
than the most massive stable isotope.  
This range of isotopes is needed to follow the nuclear
burning through hydrostatic oxygen burning.  For elements
of higher charge than zinc, the isotopes included range from the lowest mass
stable isotope to the isotope two mass units greater than the most massive
stable isotope.  This range of isotopes comfortably includes all species
important for the s-process.

The required nuclear data come from a variety of sources.
Strong and electromagnetic nuclear reaction rates are for the most
part from \cite{CF88} and \cite{tat87}.
The new reactions rates for $^{18}$O($\alpha$,$\gamma$)$^{22}$Ne and
$^{22}$Ne($\alpha$,$\gamma$)$^{26}$Mg from \cite{K94} are applied
in our calculations.
The possibility that the $^{22}$Ne($\alpha$,n)$^{25}$Mg rate might be
enhanced due to a possible resonance at 633 keV (\cite{K94}) is 
also considered by performing test calculations including the modified rate
(see \S 5).  In the Ne-Na-Mg-Al region, we use certain updated rates from
\cite{EC95}.

The rates for the neutron-capture reactions
are taken from \cite{bvw92}. These authors have
considered the dependence of these cross sections on temperature. If
the s-process occurs at high temperature, for example near kT=90 keV as
in carbon-shell burning in massive stars, then the neutron-capture cross
sections no longer follow the simple 1/v-law and the kT=30 keV
thermally-averaged cross
sections (\cite{bao87}) are no longer strictly valid. Therefore, we have
used the \cite{bvw92} cross sections to permit study of the s-process under
relatively extreme conditions.

The nuclear masses and most weak interaction (electron-capture and
$\beta$-decay) rates are taken from \cite{tuli95}. Certain weak rates are
temperature and density dependent.  For these rates we use the calculations
of \cite{tak87}. Extrapolations of some
weak interaction rates (such as  the $\beta$-decay rate of $^{79}$Se)
to higher temperatures were necessary and were done based on
experimental results (\cite{KK88} and also the data given in 
Table 2 of \cite{R93}.

%
%
\section{ CHARACTERISTICS OF THE EVOLUTIONARY MODELS}

We have evolved models of stars of masses 15, 20, 25, and 30$\msun$ with
initial solar-like composition (see \cite{AG89}) from the main 
sequence up to the ignition of central neon. The network described in
\S 3 has been coupled (between the time steps) to the stellar models
during all these evolutionary phases. In this way, we are able to determine 
correctly the neutron density during all these burning phases.
To our knowledge, no such extended calculations have been done before 
for the massive stars considered here. For example, in the recent work by
\cite{Ch98} the effect of the neutron capture reactions has 
been ignored during central helium burning, with the result that many isotopic
abundances which are affected by neutron capture reactions may not be
well determined. 

In total, we ran ten stellar models.  These are summarized in
Table \ref{tab:stellarModels}.  
Our reference calculations are case A, for which we used
a reaction rate expression for $^{12}{\rm C}(\alpha,\gamma)^{16}$O
1.7 times larger than the \cite{CF88} value (as suggested by the work
of \cite{WW93}) and the \cite{CF88} rate
expression for $^{22}{\rm Ne}(\alpha,n)^{25}$Mg.  The case B calculations
were identical to those in case A except we included a
low-energy resonance in the $^{22}{\rm Ne}(\alpha,n)^{25}$Mg reaction.  In
particular, we used the rate expression from \cite{K94} which included a
10\% contribution from this resonance.  For both case A and B we computed
models for 15, 20, 25, and 30 $\msun$ stars.  To explore the sensitivity of our
results to the $^{12}{\rm C}(\alpha,\gamma)^{16}$O reaction, we also
ran 20 $\msun$ models for the \cite{CF88} (case C) and \cite{CFHZ85} values
for the reaction rate (case D).  We find interesting consequences for both
of these rates.
 
In the following, we describe some general properties of our stellar
models.
We first present
in Fig. \ref{fig:TCvsRhoC} the evolution of the
central temperature T$_{\rm c}$ versus the central density during 
the various evolutionary phases indicated there for the four case A stars.
Calculating the stars through these phases means that we can also
follow the s-process in regions where the different shell sources are
active, especially the contribution of the
carbon-burning shell to the s-process.  The dependence of the
central temperature $T_c$ on the central density $\rho_c$ 
(in the phases before central carbon ignition) is
reasonably well approximated as 
$T_c \approx 2\times 10^7 \; (\rho_c [{\rm g \; cm}^{-3}])^{1/3}$ K,
although in general the more massive stars reach given central temperatures
at lower densities or have higher central temperatures for given
densities.
The $T_c \propto \rho_c^{1/3}$ indicates
cores of our massive star models behave approximately as polytropes
of index three, showing that the radiation pressure in the stars
is not negligible (e.g. \cite{cla83}).
After the central carbon ignition, the $T_c \propto \rho_c^{1/3}$ relation
is modified as neutrino energy loss radiates away entropy from the core 
(e.g. \cite{Arnett96}).

Table \ref{tab:evolmod}
we summarize some of the details of our four case A stellar models at key
stages in their evolution.  Our stellar models end their central helium
burning as red supergiants.  Also interesting is the increase of the helium
core mass $M_\alpha$ with time, which indicates the efficiency of accretion
from the hydrogen-shell burning.  After the end of core carbon burning or
even somewhat earlier, however, $M_\alpha$ is essentially fixed because of
the short duration of the subsequent burning phases.
An MPEG movie showing the evolution of temperatures, densities, 
luminosities, and mass fractions of $^1$H, $^4$He, $^{12}$C, $^{16}$O,
and $^{20}$Ne as a function of internal mass radius in
a 25 M$_{\sun}$ star can be viewed
at the Clemson nuclear astrophysics web site,
http://photon.phys.clemson.edu/movies.html.

Table \ref{tab:Hburnprodfact} shows the overabundances inside the hydrogen
exhausted core at the end of core hydrogen burning for the four case A
models (stage 2 in table \ref{tab:evolmod}).
Only those stable or long-lived
isotopes with overabundances noticeably
different from unity are shown.
These are the isotopes that changed during hydrogen burning.  Because the
stellar models all began with solar composition, overabundances
of unity equate to solar abundances.  Some
interesting changes have occurred during the core hydrogen burning.
$^1$H, $^3$He, and $^7$Li are all strongly depleted, as expected.  $^4$He
is enhanced by production from hydrogen.  The abundances of the CNO
isotopes have shifted around giving enhanced $^{13}$C and $^{14}$N, as
expected.  Ne-Na-Mg cycle burning alters the abundances of isotopes of
fluorine to magnesium.  The neutron source $^{22}$Ne is depleted at this
stage.  It will be created during subsequent He burning.
Finally, the abundances of $^{138}$La, $^{163}$Dy,
and $^{187}$Re all drop due to temperature enhanced weak-decay rates.  The
abundance of $^{187}$Os rises because it is the daughter of $^{187}$Re
decay.  The abundances of all other isotopes, however, are unaffected by
the core hydrogen burning.
 
The next stage of the star's evolution, and the one of primary interest for
the present paper, is core helium burning.
The evolution of each convective core mass $M_{cc}$ for
the stars under study during central helium burning is shown
versus time in Fig. \ref{fig:MconvXHe} (upper panel), and 
versus central helium mass fraction,
X($^4$He)$_{\rm c}$ (referred to as Y$_{\rm c}$ in the following) 
in Fig. \ref{fig:MconvXHe} (lower panel).
The step-like structures visible in these figures can be easily removed
by finely resolving the edge of the convective core, that is, by adding more
meshpoints there.  Our experience shows that this kind of fine resolving is
not required as long as Y$_{\rm c}$ is larger than 0.10, since the convective
core grows sufficiently slowly that a sudden uncontrolled injection of helium into the
core does not occur. 
When Y$_{\rm c}$ decreases below 0.10, however, the above stable situation  
changes drastically.  As the helium supply dwindles, the core contracts
gravitationally.  The convective core grows in response to the release of 
the gravitational binding energy.  This allows the helium-depleted core to
ingest suddenly a large amount of the overlying $^{4}$He which in turn
dramatically increases
the central energy generation rate.  This halts the central contraction and
allows the convective core to shrink again.
This behavior is called ``breathing pulses''
in the literature (e.g. \cite{cas85}).
We think that the breathing pulses are not physical.  They are rather a
numerical artifact of the necessity of discretizing
the stellar model. They 
can be avoided by careful rezoning at the edge of the convective core,
or perhaps also by coupling mixing and nucleosynthesis during the
iteration cycle, which is of course extremely time
consuming. We choose the former strategy in the present calculations with 
the result that the convective core
grows smoothly as seen in Fig. \ref{fig:MconvXHe} 
below X($^4$He)=0.10 and thus that the helium mixed into the convective core
during each time step behaves in a regular way.

To make the nuclear consequences of this point more clear, we note that
without the fine resolution of the convective core
edge but also without any limitation on the breathing pulses we would not find 
the relatively high mass fractions of $^{12}$C at the end of central
helium burning  given in Table \ref{tab:COMassfrac}.  For example, had we
not finely resolved the edge of the convective core in our case A 25 $\msun$
stellar model, the resulting central $^{12}$C and $^{16}$O mass fractions
would have been 0.17 and 0.79, respectively, in place of the corresponding
values 0.22 and 0.75 in case A.  The breathing pulses allow sudden growth
of the convective core and ingestion of
additional $^4$He and, consequently, more conversion of $^{12}$C into $^{16}$O
via the $^{12}{\rm C}(\alpha,\gamma)^{16}$O reaction.  Importantly for the
s-process, the breathing pulses also burn more $^{22}$Ne and yield a more
robust s-process (e.g. \cite{K94}).  Also consequential
is the fact that the greater
the destruction of $^{22}$Ne in core helium burning, the less efficient will
be the s-process in later burning stages, particularly shell helium and
carbon burning.

Stellar modelers typically avoid the problem of breathing pulses by ``limiting''
the growth of the convective core at low core helium mass fractions, that
is, by using computer code commands
to prevent the core from growing in size once the central helium mass
fraction falls below some value.
We ran another 25 $\msun$ calculation identical
to our case A model but without fine mass resolution of the convective core
edge and including a limitation on the growth of the core for
$Y_{\rm c} < 0.10$.
The resulting central $^{12}$C and $^{16}$O mass fractions were 0.25 and
0.73, respectively.  The limiting prevented any core growth at low central
helium and thus allowed less $^{12}$C to burn to $^{16}$O.  For the same
reason, less $^{22}$Ne burned, thereby giving a weaker s-process.

The model with suppression of the breathing pulses but without
fine mass resolution gives results reasonably close to our case A model.
Nevertheless, the case A model is clearly the more satisfactory of the
two because it does not have an artificial constraint on it.  We have therefore
chosen in our models to resolve the edge of the convective core finely since
it is most realistic scenario for our chosen convection criterion and gives
the most believable results for the s-process.

In Table \ref{tab:stellarMod} we compare the details of core helium burning
in our four case A stars with the results of other authors, where
available.  In general, our results agree well with those of \cite{P90},
who used the stellar models of \cite{N88}, both in terms of the maximum
convective core mass and the helium burning lifetime.  This makes sense
because both our models and those of \cite{N88} used the Schwarzchild
criterion for convection and employed
no overshooting.  On the other hand, both sets
of models differ from those computed with the FRANEC code (\cite{K94}).
In particular,
our maximum convective core masses are at least $\sim 10\%$ smaller than
those produced with FRANEC.  At the same time, the helium burning lifetimes
in the FRANEC models are considerably shorter than in our models and those
of \cite{P90}.  The larger cores maintain a higher supply of $^4$He and
a faster $^4$He consumption rate.  It is
interesting that the FRANEC calculations presented in \cite{K94} are rather
similar to those in \cite{S92}, which included moderate overshooting
(by 20\% of the pressure scale height).
It is eventually the treatment of mixing at the edge of the convective
core that simulates the effect of overshooting, even though it is probably
not explicitly included in the FRANEC calculations.
These stellar model differences have implications for s-process
nucleosynthesis, as discussed \S \ref{sec:sproc}.

%
%
\section{s-PROCESS DURING CENTRAL HELIUM BURNING}    \label{sec:sproc}

In this section we present and analyze our results for the s-process.  We
also compare our findings with those of other authors.

\subsection{s-Process Diagnostics}

Two quantities usefully characterize the efficiency of s-process
nucleosynthesis. One is the number of neutrons captured per iron seed:

\begin{equation}
 n_{\rm c} \equiv \sum_{A=56}^{209}(Y_A-Y_A^0)(A-56)/Y_A^0,
\label{eq:nc}
\end{equation}
where $Y_A$ is the final abundance of a nuclear species of mass number A,
and $Y_A^0$ is its initial value. 
The other quantity is the neutron exposure defined as:

\begin{equation}
  \tau_n(m_r) \equiv \int_0^t n_n(m_r,t') \; v_{\rm th} \; dt',
\label{eq:taun}
\end{equation}
where $n_n$ is the neutron density at a relative mass $m_r$, and
$v_{\rm th}=(2kT/m_n)^{1/2}$ is the thermal velocity of the neutrons.
In the present calculations, we take $v_{th}$ evaluated at $kT = 30$ keV.

It is possible to define an average neutron exposure according to the
above definition if the s-process occurs in a convective core as
during central helium burning  in massive stars, that is 

\begin{equation}
  <\tau> = \int_0^t <n_n> \; v_{\rm th} \; dt',
\end{equation}
where $<n_n>$ is the neutron density averaged over the convective core
and $v_{th}$ is the neutron thermal velocity at 30 keV.
However, such a quantity represents only a rough indicator of the efficiency
of the s-process, since the neutrons are actually locally absorbed,
mostly at the central region of the star.
A more reliable quantity is an exposure defined in terms of a nucleus like
$^{54}$Fe which is only destroyed during the s-process:
\begin{equation}
       {\rm  \tau_{54}= -{1 \over \sigma_T} \; ln (X_{54}/X_{54}^0)},
\label{eq:tau54}
\end{equation}
where $\sigma_T$ is the neutron-capture cross section at some temperature $T$
(usually taken to be 30 keV), and 
X$_{54}$, X$_{54}^0$ are the final and initial mass fractions of that nucleus
respectively.  

\subsection{Our Results}

Figs. \ref{fig:centralsproc25Msun}a-d show 
the central $\tau_{\rm n}$, $\tau_{54}$, n$_{\rm C}$,
n$_{\rm n}$ and
X$_i$/X$_{i\odot}$ versus central X($^4$He) for a 25 $\msun$ stellar model
in cases A and B.  Again, case A is the reference calculation which uses
the \cite{CF88} rate for the $^{22}{\rm Ne}(\alpha,n)^{25}$Mg reaction,
while case B includes a 10\% contribution from a
putative resonance.
According to the discussion by \cite{K94}, 
the contribution of such a resonance to this reaction rate cannot yet be
excluded on experimental grounds.

From Figs. \ref{fig:centralsproc25Msun}a-d it is clear that the
effect of the resonance would be to augment the
s-process.  For example, Fig. \ref{fig:centralsproc25Msun}b shows that
the central $^{22}$Ne abundance decreases much more dramatically in case B
than case A.  The result is a larger release of neutrons and, consequently,
a higher $n_c$ (Fig.  \ref{fig:centralsproc25Msun}c), considerably more
$^{80}$Kr and $^{25}$Mg (Fig. \ref{fig:centralsproc25Msun}b),
and a larger central $\tau_n$ (Fig. \ref{fig:centralsproc25Msun}a).
Interestingly, the sharp increase of the $^{80}$Kr 
mass fraction occurs earlier in time (or at higher Y$_{\rm c}$) in case B
than in case A due to the more efficient $^{22}$Ne destruction.
The 15, 20, and 30 $\msun$ models show qualitatively similar results;
therefore, on the basis of these results, we can conclude that  
a tight constraint on the s-process in massive stars is not possible 
as long as the the rate of $^{22}$Ne($\alpha$,n) reaction is experimentally
unsettled.

Fig. \ref{fig:centralsproc25Msun}a shows the two neutron exposures $\tau_n$
and $\tau_{54}$.  It is important to distinguish them.
The $\tau_n$ shown in Fig. \ref{fig:centralsproc25Msun}a is 
the neutron exposure at the center of the
star, that is, the $\tau_n$ computed from eq. (3) at $m_r=0$.  It
is thus the neutron exposure that would have been seen by nuclei that
always remained in the center of the star.  The value of this quantity at
the end of core helium burning is greater than about 4 $mb^{-1}$, which is an
extraordinarily large exposure.  It results from the high neutron density
in the core of the star, as shown in Fig. \ref{fig:centndnsvsXHe},
where the temperature is highest and the $^{22}$Ne
burns most effectively.  Outside the very core of the star, the temperature
is lower and the neutron density is considerably less.

A neutron exposure of 4 $mb^{-1}$ or greater is large enough to
drive all initial iron seed nuclei up to bismuth.
This, of course, is not what happens.  Because of convective mixing,
a given nucleus spends only a small fraction of
the helium-burning time in the star's center.  As mentioned above, the
neutron density outside the very central region of the star is much less,
so little s-processing is occurring there.  
This is illustrated in Fig. \ref{fig:tauneutvsMr},
which shows $\tau_n$ as a function of interior mass for our eight models.
Only the inner few
tenths of a solar mass have a significantly large $\tau_n$, and it is only
here that neutron captures are occurring.  Nuclei are convected down into
these inner regions, capture neutrons, and then mix back out into the core.
The result is that nuclei, on average, see an exposure that is much less
than the $\tau_n$ at the center.

A better measure of the average neutron exposure seen by nuclei during core
helium burning is $\tau_{54}$ at the center of the star.  It is shown in
Fig. \ref{fig:centralsproc25Msun}a and is instantaneously 
computed from eq. (5) using the current
value of the $^{54}$Fe mass fraction.  The definition of the
mass fraction of $^{54}$Fe, $X(^{54}{\rm Fe})$, is
unambiguous because $X(^{54}{\rm Fe})$ is uniform throughout the convective
zone.  $^{54}$Fe nuclei, which are only
destroyed during the s-process, are constantly
capturing neutrons in the center of the star and mixing throughout the
core; thus, $\tau_{54}$ is an appropriate measure of the average neutron
exposure.  The finding is that this average exposure is considerably less
than $\tau_n$ at the center of the star, as expected, and better
characterizes the resulting s-process abundances.  Nevertheless, even
$\tau_{54}$ is not the complete picture.  A single s-process abundance
pattern characterized by the final neutron exposure $\tau_{54}$ is
not an adequate description of the true abundance pattern at the end of
core helium burning.  In fact, a distribution of neutron exposures is
needed to characterize the final abundances.  The reasons for this are
complex, and we address them in a forthcoming paper.

To illustrate the effect of the $^{12}{\rm C}(\alpha,\gamma)^{16}$O reaction
rate on the s-process, we present Fig. \ref{fig:centralsproc20Msun}.  
This figure shows the
results for the s-process for the 20 $\msun$ star
in cases A, C, and D.  For the
lowest value for the reaction rate (case C: \cite{CF88}), the s-process is
considerably more robust than for higher values of the reaction rate.
The reason, as pointed out by \cite{R91}, is that alpha
particle capture on $^{12}$C competes with that on $^{22}$Ne for
consumption of $^4$He; thus, a lower rate of $^{12}{\rm
C}(\alpha,\gamma)^{16}$O leaves more alpha particles available for the
$^{22}{\rm Ne}(\alpha,n)^{25}$Mg reaction and consequently leads to
more neutron production.  It aids understanding to recall that most of the
$^{22}$Ne consumption occurs when less than 10\% of the helium remains
(Fig. \ref{fig:centralsproc25Msun}b and \ref{fig:centralsproc20Msun}b).

The faster $^{12}{\rm C}(\alpha,\gamma)^{16}$O reaction rate explains the
less efficient s-process in case A than in cases C and D.  
Counter to expectations, however, case D shows a more efficient
s-process than case A, even though the $^{12}{\rm C}(\alpha,\gamma)^{16}$O
reaction rate is always greater in case D over the temperature range
of interest for the core helium.  The resolution 
of this puzzle lies in the fact that the faster 
$^{12}{\rm C}(\alpha,\gamma)^{16}$O reaction rate 
(Case D rate is $\simeq$2.4 $\times$ rate of Case C at temperature
$\sim$2.5$\times$10$^8$ K)
also leaves fewer
$^{12}$C nuclei around ($^{12}$C mass fraction at the end of
core He burning is 0.12 in Case D and 0.23 in Case C).
The smaller abundance of $^{12}$C can compensate
the faster rate leading to a larger abundance of alpha particles and
greater consumption of $^{22}$Ne.  When the triple-alpha reaction is no
longer the major consumer of alpha particles (owing to the small value
of $Y_c$), the $^{12}$C and $^{22}$Ne compete in proportion to their
abundances for the final helium.  In fact, careful examination of 
Fig. \ref{fig:centralsproc20Msun} shows that the various curves behave in
strict accordance with the rule that a larger
$^{12}{\rm C}(\alpha,\gamma)^{16}$O reaction cross section
gives less s-processing
until the $^4$He mass fraction drops to about 0.01.  Then cases A and D
cross over.  It is here that the lower abundance of $^{12}$C in case D
overcomes the larger reaction cross section.

The same interplay between $^{12}$C and $^{22}$Ne explains the difference
between the case A and B central $^{12}$C mass fractions in table
\ref{tab:COMassfrac}.  One would expect a faster $^{22}{\rm Ne}(\alpha,n)
^{25}$Mg rate (case B) to allow $^{22}$Ne to compete more effectively for
alpha particles, thereby resulting in a higher final $^{12}$C mass fraction.
In fact, the opposite is true.  The case B $^{12}$C mass fractions are
consistently slightly
{\em lower} in table \ref{tab:COMassfrac} than the corresponding
case A values.  The reason is that so much $^{22}$Ne burns in the case B
models that its low abundance overcomes the faster rate and makes it compete
for alpha particles slightly less effectively than in case A.

\subsection{Comparison of s-process Studies}

In Table \ref{tab:sproccompareCF88} 
we present a comparison of our results for the
s-process at helium exhaustion and those obtained by \cite{P90} and
\cite{K94} when all three sets of calculations used the \cite{CF88}
expression for the $^{22}{\rm Ne}(\alpha,n)^{25}$Mg reaction rate.
Before discussing the differences, it is useful to point out
two consistent trends within the results of each set of authors.
First, the more massive the star,
the more $^{22}$Ne is burned.  This reflects the higher
core temperatures in more massive stars and, consequently, the higher rate
of alpha particle capture on $^{22}$Ne.   Second, it is consistently the
case that the lower the $^{22}$Ne abundance at core helium exhaustion, the
greater the number of neutrons captured per iron seed nucleus.  This is
simply due to the fact that the neutrons driving the s-process are
liberated from $^{22}$Ne.

Now we point out the differences between our results and those
of the other authors.  First, (using our reference calculation case A), our
models 
typically burn less $^{22}$Ne than \cite{P90} but more than \cite{K94}.
Second, we consistently get more s-processing than either \cite{P90} or
\cite{K94}.

Our models typically burn less $^{22}$Ne than \cite{P90} because these latter
authors used the \cite{CF88} rate value for the $^{12}{\rm
C}(\alpha,\gamma)^{16}$O reaction.  As with our case C model discussed in
the previous subsection, a low value for this rate means more alpha particles
are available for capture on $^{22}$Ne during the s-processing phase.  

Despite the fact that our
models typically burn less $^{22}$Ne than those of \cite{P90},
our models in fact show a larger number of
neutrons captured per iron seed nucleus.  We believe the reason for this is
that our calculations used temperature-dependent neutron capture cross
sections (\cite{bvw92}) while \cite{P90} used the 30 keV cross sections of
\cite{bao87}.  Much of the s-processing in our models happens at
temperatures in the range of about 23 to 26 keV, and the cross sections for
typical s-process nuclei in
this temperature range can be up to factors of two larger than at 30 keV.
On the other hand, the cross sections for major s-process poisons
at 23 to 26 keV can be slightly lower than at 30 keV.
Our models therefore
also show greater s-processing because of more efficient neutron capture
by heavy nuclei.
It is of interest to note that the rather large
differences in the capture cross sections as a function of temperature
probably reflect the role of resonances in the compound nuclei.

Finally, it is worth pointing out that our case C 20 $\msun$ stellar model
leaves less $^{22}$Ne than the corresponding \cite{P90} model even though
they both used the same $^{12}{\rm C}(\alpha,\gamma)^{16}$O reaction rate
value.  Our best explanation for this is some difference in the stellar
models, perhaps in differing convective histories.  This is plausible since
there are rather strong differences in $<\tau >$ between our models and
those of \cite{P90}.  Interestingly, our case
D 20 $\msun$ model burns about the same amount of $^{22}$Ne as does the
corresponding \cite{P90} 20 $\msun$ model.  Our model shows more
s-processing, however, which we again attribute to somewhat higher
neutron capture cross sections in our models during the s-process phase.

The models of \cite{K94} show consistently less s-processing than either
our models or those of \cite{P90}.  This is true despite the fact that our
models compare rather favorably to those of \cite{K94} in terms of $<\tau
>$ and maximum neutron density averaged over the convective core.  These
differences are also present despite the fact that the neutron capture cross
sections used in the two sets of calculations are the same and the
$^{12}{\rm C}(\alpha,\gamma)^{16}$O reaction rates do not differ greatly.  
To explain the differences, therefore, we again appeal to differences 
in stellar models.  
Table \ref{tab:stellarMod} compares the maximum convective core masses 
and helium burning durations of the various models.  
Our models compare quite well with those of \cite{P90}.  
On the other hand, the models of the FRANEC code
used in \cite{K94} have consistently larger maximum convective cores and
shorter helium burning lifetimes.  Interestingly, the convective cores and
helium burning durations of the \cite{K94} models are closer to those of
\cite{S92} which included convective overshooting during core helium
burning.  The treatment of mixing in the FRANEC models presented by \cite{K94}
leads to enlarged convective core masses similar to the overshooting case.
Whatever the reason for the
differences in the stellar models, the shorter helium burning lifetimes in
the \cite{K94} allowed for less s-processing compared to our models and
those of \cite{P90}.

In Table \ref{tab:sproccompareK94} we present a comparison of s-process
results when the tentative $^{22}{\rm Ne}(\alpha,n)^{25}$Mg rate of
\cite{K94} is adopted.  The first obvious conclusion is that the
s-processing is considerably stronger both in our models and those of
\cite{K94} because of the larger rate for alpha capture on $^{22}$Ne.
We consistently find a larger number of neutron captures per iron seed
nucleus, however, again most likely due to differences in stellar models,
as discussed above.  It is interesting that we find smaller overproductions
of $^{80}$Kr in the 25 and 30 $\msun$ stellar models than do \cite{K94}.
This is due to the fact that the s-process in our models is so strong that
enough neutron capture has occurred to push the peak overproductions past
Kr and into the Sr region (see the next subsection).
For the same stellar masses, the models of
\cite{K94} have overproductions still peaking at $^{80}$Kr.

\subsection{s-Process Production Factors}
The production factors at the end of central helium burning are
displayed in Fig. \ref{fig:prodfactRCF88} 
and \ref{fig:prodfact25} and the numerical values are presented in Table
\ref{tab:overprod}.
The results show the well known feature of the weak
s-process component that is the synthesis of the heavy elements in
the mass range A=60-90, with the largest production factor for
$^{80}$Kr. 

An interesting result is that in our models with the
$^{22}$Ne($\alpha$,n)$^{25}$Mg rate of \cite{K94} (case B), $^{80}$Kr is
no longer the most overproduced isotoped in core helium burning in the 20,
25, and 30 $\msun$ stellar models.  $^{86}$Sr or $^{87}$Sr show larger
overproductions.  These isotopes are factors of two or three less
overproduced when the $^{22}$Ne($\alpha$,n)$^{25}$Mg rate of \cite{CF88} is
used.

Apart from the large overabundances of the weak s-process nuclei, table
\ref{tab:overprod} shows significant production of several interesting
isotopes.  $^{12}$C and $^{16}$O are produced as
the main products of helium burning. Considerable $^{22}$Ne is present because
it was left unburnt.  $^{25,26}$Mg are produced via alpha particle
capture on $^{22}$Ne.  A number of isotopes in the mass range 30-60 are
made by neutron capture.  Chief among these is $^{40}$K.  Above mass number
104, the only significant production is of $^{152}$Gd, produced by branching
at $^{151}$Sm and $^{152}$Eu, and of $^{180}$Ta.  This latter isotope is
made in the mechanism first described by \cite{YT83}.  The idea is that
$^{179}$Ta is unstable to electron capture
in the star, but temperature-enhanced $\beta^-$ decay from $^{179}$Hf
in the helium-burning
core allows a steady-state abundance to build up and neutron capture to
$^{180}$Ta.  We mention that in our current models we allow no equilibration
between the $^{180}$Ta ground state (which decays to $^{180}$Hf in 8.15 hours)
and the metastable state at 75 keV (which is essentially stable).
The large spin difference between the ground and metastable states prevents
direct transitions between them; thus, equilibration would have to occur
through higher-lying nuclear levels.  This
could occur at the temperatures present in core helium burning
through intermediate states at energies in the range of 500 - 1000 keV.
Were such states indeed present, the metastable state could depopulate,
thereby resulting in less $^{180}$Ta production.  The experimental situation
concerning the existence of such states is not yet clear.  One group
finds evidence for the existence of at least one such state (\cite{Sch94};
\cite{Neu97}) while another does not (\cite{Schu97}).  A firm prediction
of $^{180}$Ta production in any s-process environment
will require clarification of the nuclear physics in this interesting
region of the chart of the nuclides.

A final comment concerns the mainstream presolar SiC grains, which, having
condensed in the outflow from AGB stars, the site of the main component,
should contain differences from solar abundances that compensate for the
portions produced in massive stars.  An interesting puzzle is the fact that
the Sr isotopes are almost normal in SiC grains (\cite{Nic98}; \cite{Pod98}),
whereas our calculations of the massive-star component show large ratios
for $^{86}$Sr/$^{88}$Sr and $^{87}$Sr/$^{88}$Sr.  This imbalance ultimately
requires explanation either through Galactic abundance evolution of the Sr 
isotopes or perhaps their modification in more advanced stellar burning phases.

\section{The s-Process in Advanced Burning Phases: A Preview}
\label{sec:advanced}

Though our primary purpose in this paper has been to study the s-process during
core helium burning, we briefly present in this section some results of the
s-process in advanced burning phases as a preview to a more extended study
to appear 
in a forthcoming paper.  We summarize some of the most interesting results
for the 25 $\msun$ model in Fig. \ref{fig:kr80Cshell}. 
The upper panel of this figure shows $\tau_n$
and $\tau_{54}$ as functions of interior mass at the end of core helium
burning (dashed lines) and near the end of core neon burning (solid lines).
The peaks in $\tau_n$ at $m_r \approx 1.6 \msun$ and $m_r \approx 5.8
\msun$ indicate regions of strong neutron capture in the carbon burning and
helium burning shells, respectively.  The s-processing during these phases,
resulting largely from burning of residual $^{22}$Ne, increases $\tau_{54}$
throughout the star over its value at the end of core helium burning.

The s-processing occurring during these shell burning phases has a dramatic
effect on the overabundance of $^{80}$Kr.  As seen in the middle panel of
the figure, core and shell carbon burning deplete by factors of ten or more
the huge $^{80}$Kr overabundance left from core helium burning.  The
destruction occurs because of the high neutron densities (up to several
times $10^{10}$ cm$^{-3}$) present in the
innermost part of the carbon-burning shell.
Because of the high neutron density in this zone,
$^{79}$Se captures a neutron to $^{80}$Se faster than it beta decays.
This in turn allows the s-process flow for the most part to bypass $^{80}$Kr,
thereby depleting its previously large
abundance.  The intense energy release due to the
carbon burning in this zone drives convection out to 5 $\msun$; therefore,
nuclei mix down into the high-neutron-density zone, capture neutrons, and
mix back out into the outer shell.  This mixing and burning depletes the
$^{80}$Kr throughout the convective shell.  Significantly, $^{22}$Ne is also
continually mixing down into the burning zone.  This replenishes the
neutron source and keeps the neutron density high for a much longer time
than would be the case without mixing.  
An MPEG movie illustrating the evolution of the overabundances of
$^{80}$Kr, $^{22}$Ne, $^{13}$C, $^4$He and neutron density in the 
25 M$_{\sun}$ stellar model is available for viewing on the world-wide web at
http://photon.phys.clemson.edu/movies.html.

It is interesting that destruction of $^{80}$Kr constrains the mixing in
the carbon shell to be very rapid. The lifetime of a $^{79}$Se nucleus
against beta decay is only
about 0.4 years in the innermost part of the shell at 1.6 $\msun$
(where the temperature is about $10^9$K) and about 4 years in the outermost
part of the shell at 5 $\msun$
(where the temperature is about $3 \times 10^8$K).  In order to ensure that
$^{79}$Se nuclei throughout the shell can be re-exposed to neutrons before
they beta decay and that $^{80}$Kr in fact gets bypassed, the timescale for
overturn of the carbon shell must be less than several tenths of a year.  Our
stellar models give such rapid mixing.  The convective
velocities in the carbon-burning shell are of the order of 1 $km/s$ while
the physical size of the shell is $\sim 40,000$ km thick.  The shell overturns
once every $\sim 10^5\ s$ or $0.03\ yr$, sufficiently fast to re-expose
the $^{79}$Se nuclei to the intense flux.

Interestingly,
$^{80}$Kr is also destroyed outside about 5.2 $\msun$ due to shell helium
burning, which yields neutron densities as high as a few times $10^{10}$
cm$^{-3}$ at times earlier than that shown here.
The neutron density is comparable to that in the carbon burning shell.
This is somewhat surprising because the temperature and thus the
the $^{22}{\rm Ne} (\alpha, n)^{25}{\rm Mg}$ cross section are both lower
in the helium burning shell.  On the other hand, the alpha particle
abundance is considerably higher, allowing for efficient production of
neutrons.  Because of the efficient burning in the He and C shells,
large $^{80}$Kr overproductions from core helium burning
only persist in a thin region
between the outer edge of the convective carbon-burning shell and the inner
edge of the helium-burning shell in our models.
The lower panel of the figure shows the
abundances of the major species at the given moment in the star's life.

In summary, our results
indicate that $^{80}$Kr, vastly overproduced in core helium burning, can be
subsequently largely destroyed in advanced burning phases.  The reduced yield
of $^{80}$Kr due to destruction during advanced burning phases has strong
implications for the chemical evolution of that isotope.
We intend to analyze this question in much more detail in a
forthcoming paper.

\include{conclusions}

\acknowledgments
  
We are grateful to Donald Clayton for helpful comments and suggestions.
This work has been supported by NASA grants NAG5-4329, NAGW-3277, and
NAGW-3480 and by a UR grant (5-30-1911-42-0029) at Clemson University.


\include{bibitem}
\clearpage


\include{tables}



\include{figures}
\clearpage

\end{document}

%% file: definition.tex
\def\anj#1,#2.{{\sl Astron. J.} {\bf #1}, #2.}

\def\zfa#1,#2.{{\sl Z. Astrophys.} {\bf #1}, #2.}
\def\mnras#1,#2.{{\sl Monthly Notices Roy. Astron. Soc.} {\bf #1}, #2.}
\def\msun{\, {{\rm M_{\odot}}}}       

\def\beqa{\begingroup{\global\advance\eqnuma by 1}$$}   

\def\chap#1{ \global \advance\chapnum by 1  \chpreset \chskipt
             \line{ \the\chapnum . \chpindent #1\hfill}
             \penalty 100000 \chskipl \penalty 100000 }
\def\:{\mskip\medmuskip}                         
 \def\sqr#1#2{{\vcenter{\hrule height.#2pt
     \hbox{\vrule width.#2pt height#1pt \kern#1pt
           \vrule width.#2pt}
           \hrule height.#2pt}}}
 

%% file: conclusions.tex
\section{CONCLUSIONS}

We have performed detailed calculations of the s-process in massive stars
with up-to-date neutron capture cross sections (\cite{bvw92}),
temperature and density-dependent weak interaction rates (\cite{tak87};
\cite{KK88}), 
and a full s-process reaction network coupled to the stellar model.
Consequently, we can determine the composition after core helium burning
phase more accurately than calculations using post-processing.
In particular, we are able to determine the dependence of the
neutron exposure ($\tau$) and neutron capture per iron seed nucleus 
($n_{\rm C}$) on mass radius at each time step, especially for the
advanced evolution phases where average values are no longer possible
or relevant.

From the results presented in this paper, we conclude
\begin{enumerate}

\item Fine rezoning of the the mesh zones near the edge of convective 
core can avoid the so-called core breathing problem during helium burning.
The convective cores in our models grow and shrink smoothly;
therefore, our treatment of the convective mixing is accurate
within the framework of the Schwarzchild criterion.

\item A possible low-energy 633 keV
resonance in the $^{22}$Ne($\alpha$,n)$^{25}$Mg reaction, which would
dramatically increase the rate over the \cite{CF88} value, will significantly
enhance the robustness of the s-process in massive stars.  
This confirms the results of \cite{K94}.  
Accurate constraints from massive star s-processing must therefore 
await experimental resolution of the question of the existence of 
this resonance.

\item The $^{12}$C($\alpha$,$\gamma$)$^{16}$O reaction has an important 
effect on the efficiency of the s-process in massive stars.  
In particular, a low value for the rate of this reaction (such as
that given in the compilation of \cite{CF88}) leads to a more robust
s-process than a high value (such as that of \cite{CFHZ85}).  
The reason is that the slower this rate is, the more alpha
particles are available for capture on $^{22}$Ne.  
Nevertheless, two popular parameterizations for 
this rate (the \cite{CFHZ85} rate and 
1.7 times the \cite{CF88} rate) give similar s-process results 
to within $\sim$10-20\%.

\item Neutron captures in the core-helium-burning s-process occur only in
the most central regions of the massive star because this is where
$^{22}$Ne is burning and releasing neutrons.  For this reason, use of
$<\tau>$, the average exposure over the convective core is not a
particularly useful characterization of the efficiency of the s-process.  
A much more useful quantity is $\tau_{54}$ which accounts for 
highly-concentrated burning in the
center of the star and for convective mixing.  Nevertheless, even this
quantity is limited in its usefulness because there is in fact a
distribution of neutron exposures.  We will address this issue in a
forthcoming paper.

\item The massive star s-process is highly sensitive to the size of the
convective core and the treatment of mixing at its edge.  For example,
extra mixing (such as overshooting) can dramatically change the duration of
the helium burning phase and, consequently, the efficiency of the s-process
(see \S \ref{sec:sproc}).  Once nuclear physics issues such as the correct value of
the $^{22}$Ne$(\alpha,n)^{25}$Mg reaction rate are settled, the s-process
will be a powerful diagnostic of convection in massive stars.

\item The high $^{80}$Kr overproduction built up during core helium burning
may subsequently be strongly destroyed during carbon shell burning
(\S \ref{sec:advanced})).  We will
explore this possibility in a forthcoming paper.

\end{enumerate}

%% file: tables.tex
\begin{deluxetable}{ccccccccccccccc} 
\footnotesize 
\tablecaption{Nuclei in the Network} 
\tablewidth{0pt} 
\tablehead{ 
\colhead{Z}  & \colhead{$A_{min}$}  & \colhead{$A_{max}$} &   &
\colhead{Z}  & \colhead{$A_{min}$}  & \colhead{$A_{max}$} &   &
\colhead{Z}  & \colhead{$A_{min}$}  & \colhead{$A_{max}$} &   &
\colhead{Z}  & \colhead{$A_{min}$}  & \colhead{$A_{max}$} 
}
\startdata 
1  & 2 & 2  & & 	22 & 42 & 51 & &	43 & 97 & 101 &&	64 & 152 & 161 \\
2  & 3 & 4  & &	23 & 44 & 52 & &	44 & 96 & 105 &&	65 & 156 & 162 \\
3  & 7 & 7  & & 24 & 46 & 55 &&45 & 103 & 105 && 	66 & 156 & 165 \\
4  & 7 & 7  & &25 & 48 & 56 & &46 & 102 & 111 && 	67 & 162 & 166 \\
5  & 8 & 11 & &26 & 50 & 61 & &47 & 107 & 111 &&	68 & 162 & 171 \\
6  & 11 & 15 & &27 & 51 & 61 &&	48 & 106 & 117 &&69 & 168 & 172 \\
7  & 13 & 16 & &28 & 54 & 65 &&	49 & 103 & 117 &&70 & 168 & 177 \\
8  & 15 & 19 &&29 & 56 & 66 & &50 & 112 & 127 &&71 & 174 & 180 \\
9  & 17 & 20 & &30 & 59 & 71 &&	51 & 121 & 127 &&72 & 174 & 183 \\
10 & 20 & 23 & &31 & 69 & 72 &&52 & 120 & 131 &&73 & 179 & 182 \\
11 & 21 & 24 & &32 & 70 & 77 &&	53 & 125 & 131 &&74 & 180 & 188 \\
12 & 23 & 27 & &33 & 75 & 77 &&	54 & 124 & 137 &&75 & 184 & 189 \\
13 & 25 & 28 & &34 & 74 & 83 &&	55 & 131 & 137 &&76 & 184 & 195 \\
14 & 26 & 31 & &35 & 79 & 83 &&56 & 130 & 139 &&77 & 190 & 195 \\
15 & 27 & 34 & &36 & 78 & 87 &&57 & 137 & 140 &&78 & 190 & 199 \\
16 & 30 & 37 & &37 & 85 & 88 &&58 & 136 & 144 &&79 & 195 & 200 \\
17 & 33 & 38 & &38 & 84 & 91 &&59 & 141 & 143 &&80 & 196 & 205 \\
18 & 34 & 41 & &39 & 89 & 91 &&60 & 142 & 151 &&81 & 203 & 206 \\
19 & 37 & 42 & &40 & 90 & 97 &&61 & 144 & 151 &&82 & 204 & 209 \\
20 & 38 & 49 & &41 & 93 & 97 &&62 & 144 & 155 &&83 & 209 & 210 \\
21 & 40 & 49 & & 42 & 92 & 101 &&	63 & 151 & 157 && 84 & 210 & 254 \\
\enddata 
\tablenotetext{}{
    Z is the proton number and A is the mass number. }
\label{tab:tabnet} 
\end{deluxetable}

\begin{deluxetable}{clll}
\footnotesize
\tablecolumns{4}
\tablecaption{List of Stellar Models Studied}
\tablewidth{0pc}
\tablehead{
\colhead{Case} & \colhead{$^{12}$C($\alpha$,$\gamma$)$^{16}$O} &
                 \colhead{$^{22}$Ne($\alpha$,n)$^{25}$Mg} & 
                                        Star Mass (M$_{\sun}$) 
}
\startdata
A              & CF88 $\times$ 1.7  & CF88    & 15, 20, 25, 30 \\
B              & CF88 $\times$ 1.7  & WT K94  & 15, 20, 25, 30 \\
C              & CF88               & CF88    & 20 \\
D              & CFHZ85               & CF88    & 20 
\enddata
\label{tab:stellarModels}
\end{deluxetable}
\clearpage

\begin{deluxetable}{cccccccccc}
\tablecolumns{10}
\tablewidth{0pc}
\tablecaption{
CHARACTERISTICS OF EVOLUTIONARY MODELS}
\tablehead{ 
\colhead{stage}  & \colhead{Evol. Time}  & 
\colhead{log(L/L$_{\sun}$)} & \colhead{log(T$_{\rm eff}$)} & \colhead{T$_{c}$}&
\colhead{$\rho_{\rm c}$}& \colhead{M$_{\alpha}$}& \colhead{M$_{\rm co}$} &
\colhead{M$_{\rm NeMg}$} & \colhead{M$_{\rm cc}^{max}$} \\
  & \colhead{(yrs)}  & \colhead{} & \colhead{(K)} &
    \colhead{($10^8$ K)} & \colhead{(g cm$^{-3}$)}& \colhead{(M$_{\sun}$)} &
    \colhead{(M$_{\sun}$)} & \colhead{(M$_{\sun}$)} & \colhead{(M$_{\sun}$)} 
}
\startdata
\multicolumn{10}{c}{\underline{15 M$_{\sun}$} }  \\
1 & 0.00                 & 4.274 & 4.484 & 0.340 & 6.13$\times$10$^0$ & 
                            0.00 & 0.00  & 0.00& 5.58 \\
2 & +1.06$\times$10$^7$ & 4.617 & 4.405 & 0.628 & 6.30$\times$10$^1$ & 
                           2.30  & 0.00  & 0.00& 0.00 \\
3 & +3.30$\times$10$^4$  & 4.641 & 4.186 & 1.30  & 1.31$\times$10$^3$ & 
                           2.69  & 0.00  & 0.00 & 2.17 \\ 
4 & +1.68$\times$10$^6$  & 4.733 & 3.560 & 3.12  & 6.31$\times$10$^3$ & 
                            3.90 & 2.07  & 0.00 & 0.00 \\
5 & +3.14$\times$10$^4$  & 4.816 & 3.550 & 5.46 & 1.10$\times$10$^5$ &
                            3.91 & 2.12  & 0.00 & 0.51      \\
6 & +8.12$\times$10$^3$  & 4.849 & 3.548 & 9.81 & 5.54$\times$10$^6$ &
                            3.91 & 2.21  & 1.12 & 0.00     \\
7 & +4.85$\times$10$^1$  & 4.845 & 3.548 & 12.11 & 1.35$\times$10$^7$ &
                            3.91 & 2.24  & 1.47 & 0.00      \\
     \multicolumn{10}{c}{} \\
\multicolumn{10}{c}{\underline{20 M$_{\sun}$} } \\
1 & 0.00                & 4.625 & 4.539 & 0.359 & 4.87$\times$10$^0$ & 
                          0.00  & 0.00  & 0.00  & 8.71 \\
2 & +7.64$\times$10$^6$ & 4.982 & 4.433 & 0.751 & 6.16$\times$10$^1$ &
                          4.15  & 0.00  & 0.00  & 0.00 \\
3 & +1.38$\times$10$^3$ & 5.018 & 4.265 & 1.36  & 6.29$\times$10$^2$ &
                         4.35  & 0.00  & 0.00  & 3.79 \\
4 & +1.03$\times$10$^6$ & 5.049 & 3.551 & 3.27  & 4.37$\times$10$^3$ &
                          5.90  & 3.75  & 0.00  & 0.00 \\
5 & +1.54$\times$10$^4$ & 5.134 & 3.541 & 5.92  & 5.27$\times$10$^4$ &
                          5.90  & 3.81  & 0.00  & 0.47 \\
6 & +2.90$\times$10$^3$  & 5.155 & 3.539 & 10.350 & 3.03$\times$10$^6$ &
                          5.90  & 3.81  & 1.15 & 0.00     \\
7 & +3.81$\times$10$^1$  & 5.152 & 3.539 & 12.126 & 7.09$\times$10$^6$ &
                          5.90  & 3.81  & 1.78 & 0.00      \\
     \multicolumn{10}{c}{} \\
\multicolumn{10}{c}{\underline{25 M$_{\sun}$} }  \\
1 & 0.00                & 4.877 & 4.575 & 0.372 & 4.06$\times$10$^0$ &
                          0.00  &  0.00 & 0.00  & 12.26 \\ 
2 & +6.19$\times$10$^6$ & 5.236 & 4.395 & 0.972 & 1.07$\times$10$^2$ &
                          6.11  &  0.00 & 0.00  & 0.00 \\
3 & +5.40$\times$10$^3$ & 5.258 & 4.267 & 1.40  & 3.91$\times$10$^2$ &
                          6.36  &  0.00 & 0.00  & 5.65 \\
4 & +7.65$\times$10$^5$ & 5.279 & 3.079 & 3.61  & 4.33$\times$10$^3$ &
                          7.95  & 5.49  & 0.00  & 0.00 \\
5 & +9.40$\times$10$^3$ & 5.341 & 3.535 & 6.24  & 3.79$\times$10$^4$ &
                          7.95  & 5.52  & 0.00  & 0.30 \\
6 & +1.30$\times$10$^3$ & 5.358 & 3.533 & 12.07  & 1.63$\times$10$^6$ &
                          7.95  & 5.54  & 1.56  & 0.00 \\
7 & +5.25$\times$10$^{-1}$ & 5.358 & 3.533 & 12.18  & 1.82$\times$10$^6$ &
                          7.95  & 5.54  & 1.56  & 0.00 \\
\tablebreak                    
     \multicolumn{10}{c}{} \\
\multicolumn{10}{c}{\underline{30 M$_{\sun}$} } \\
1 & 0.00                & 5.070 & 4.601 & 0.381 & 3.54$\times$10$^0$ &
                          0.00  &  0.00 & 0.00  & 16.06 \\
2 & +5.32$\times$10$^6$ & 5.428 & 4.263 & 1.36  & 2.45$\times$10$^2$ &
                           8.51 &  0.00 &  0.00 & 0.00 \\
3 & +5.00$\times$10$^2$ & 5.428 & 4.233 & 1.44  & 2.97$\times$10$^2$ &
                           8.51 &  0.00 &  0.00 & 7.65\\
4 & +6.17$\times$10$^5$ & 5.439 & 3.557 & 3.43  & 2.94$\times$10$^3$ &
                          10.06 & 7.55  & 0.00  & 0.00 \\
5 & +6.52$\times$10$^3$ & 5.439 & 3.550 & 6.34  & 2.71$\times$10$^4$ &
                         10.06  & 7.59  & 0.00  & 0.00 \\
6 & +7.90$\times$10$^2$ & 5.505 & 3.550 & 13.11  & 1.21$\times$10$^6$ &
                         10.06  & 7.59  & 2.02  & 0.00 \\
7 & +0.00$\times$10$^0$ & 5.505 & 3.550 & 13.11  & 1.21$\times$10$^6$ &
                         10.06  & 7.59  & 2.02  & 0.00 \\
\enddata
\tablenotetext{a}{M$_{cc}^{max}$ is the maximum convective core mass of
that burning phase.}
\tablenotetext{b}{M$_{\alpha}$ is the mass size of helium core.}
\tablenotetext{c}{T$_{C}$ is the central temperature.}
\tablenotetext{d}{$\rho_{C}$ is the central density.}
\tablenotetext{}{Stages 1 to 7 correspond to the evolutionary time of
1) Zero Age Main sequence, 2) central hydrogen exhaustion, 
3) central helium ignition, 4) central helium exhaustion, 
5) central carbon ignition, 6) central carbon exhaustion,
and 7) central neon ignition, respectively.}
\label{tab:evolmod}
\end{deluxetable}

\begin{deluxetable}{rllll} 
\footnotesize 
\tablecaption{Production factors inside hydrogen exhausted core at the end of
core hydrogen burning}
\tablewidth{0pt} 
\tablehead{ 
\colhead{isotope}  & \multicolumn{4}{c}{Production factors} \\
                   & \colhead{15 M$_{\sun}$}  & \colhead{20 M$_{\sun}$} &   
                     \colhead{25 M$_{\sun}$}  & \colhead{30 M$_{\sun}$} 
}
\startdata 
$^{1}$H    & 1.0$\times$10$^{-4}$  &   7.6$\times$10$^{-4}$   &
             4.9$\times$10$^{-4}$  &   9.3$\times$10$^{-4}$ \\
$^{3}$He   & 3.4$\times$10$^{-12}$ &   8.3$\times$10$^{-11}$  &
             4.2$\times$10$^{-11}$ &   1.1$\times$10$^{-9}$  \\
$^{4}$He   & 3.6$\times$10$^{0}$   &   3.6$\times$10$^{0}$    &
             3.6$\times$10$^{0}$   &   3.6$\times$10$^{0}$  \\
$^{7}$Li   & 2.4$\times$10$^{-13}$ &   2.1$\times$10$^{-12}$  &
             1.3$\times$10$^{-12}$ &   2.6$\times$10$^{-12}$  \\
$^{12}$C   & 8.9$\times$10$^{-2}$  &   9.3$\times$10$^{-2}$   &
             9.8$\times$10$^{-2}$  &   1.0$\times$10$^{-2}$  \\
$^{13}$C   & 2.3$\times$10$^{0}$   &   2.4$\times$10$^{0}$  &
             2.5$\times$10$^{0}$   &   2.6$\times$10$^{0}$  \\
$^{14}$N   & 1.1$\times$10$^{1}$   &   1.1$\times$10$^{1}$   &
             1.1$\times$10$^{1}$   &   1.1$\times$10$^{1}$  \\
$^{15}$N   & 9.0$\times$10$^{-2}$  &   9.3$\times$10$^{-2}$  &
             9.1$\times$10$^{-2}$  &   9.2$\times$10$^{-2}$ \\
$^{16}$O   & 3.2$\times$10$^{-2}$  &   3.0$\times$10$^{-2}$  &
             2.9$\times$10$^{-2}$  &   2.7$\times$10$^{-2}$  \\
$^{17}$O   & 9.7$\times$10$^{-2}$  &   9.1$\times$10$^{-2}$  &
             9.0$\times$10$^{-2}$  &   8.7$\times$10$^{-2}$ \\
$^{18}$O   & 5.7$\times$10$^{-6}$  &   5.4$\times$10$^{-6}$  &
             4.8$\times$10$^{-6}$  &   4.5$\times$10$^{-6}$  \\
$^{19}$F   & 2.6$\times$10$^{-5}$  &   2.6$\times$10$^{-5}$ &
             3.0$\times$10$^{-5}$  &   3.1$\times$10$^{-5}$  \\
$^{20}$Ne  & 9.9$\times$10$^{-1}$  &   9.9$\times$10$^{-1}$  &
             9.8$\times$10$^{-1}$  &   9.8$\times$10$^{-1}$  \\
$^{21}$Ne  & 1.9$\times$10$^{0}$   &   7.0$\times$10$^{-1}$   &
             2.1$\times$10$^{-1}$  &   1.1$\times$10$^{-1}$   \\
$^{22}$Ne  & 2.6$\times$10$^{-2}$  &   1.7$\times$10$^{-2}$ &
             9.8$\times$10$^{-3}$  &   6.1$\times$10$^{-3}$  \\
$^{23}$Na  & 5.5$\times$10$^{0}$   &   5.9$\times$10$^{0}$  &
             6.0$\times$10$^{0}$   &   6.1$\times$10$^{0}$  \\
$^{25}$Mg  & 6.0$\times$10$^{-2}$  &   4.4$\times$10$^{-2}$ &
             3.7$\times$10$^{-2}$  &   3.2$\times$10$^{-2}$ \\
$^{26}$Mg  & 1.8$\times$10$^{0}$   &   1.8$\times$10$^{0}$  &
             1.8$\times$10$^{0}$   &   1.8$\times$10$^{0}$  \\
$^{138}$La & 9.5$\times$10$^{-1}$  &   9.6$\times$10$^{-1}$  &
             9.6$\times$10$^{-1}$  &   9.6$\times$10$^{-1}$  \\
$^{163}$Dy & 6.4$\times$10$^{-1}$  &   7.9$\times$10$^{-1}$  &
             7.0$\times$10$^{-1}$  &   6.9$\times$10$^{-1}$ \\
$^{187}$Re & 3.6$\times$10$^{-2}$  &   1.0$\times$10$^{-2}$  &
             1.6$\times$10$^{-2}$  &   1.6$\times$10$^{-2}$ \\
$^{187}$Os & 5.2$\times$10$^{0}$   &   5.3$\times$10$^{0}$   &  
             5.3$\times$10$^{0}$   &   5.3$\times$10$^{0}$  \\
\enddata 
\label{tab:Hburnprodfact} 
\end{deluxetable}
\clearpage

\begin{deluxetable}{lll}
\footnotesize
\tablecolumns{4}
\tablecaption{Central $^{12}$C and $^{16}$O Mass Fractions at Core Helium
Exhaustion}
\tablewidth{0pc}
\tablehead{
\colhead{Model} & \colhead{$X(^{12}{\rm C})$} &
                 \colhead{$X(^{16}{\rm O})$} \\
}
\startdata
15 $\msun$, case A & 0.251 & 0.723 \\
15 $\msun$, case B & 0.219 & 0.753 \\
20 $\msun$, case A & 0.231 & 0.741 \\
20 $\msun$, case B & 0.218 & 0.752 \\
20 $\msun$, case C & 0.335 & 0.637 \\
20 $\msun$, case D & 0.115 & 0.856 \\
25 $\msun$, case A & 0.220 & 0.750 \\
25 $\msun$, case B & 0.211 & 0.757 \\
30 $\msun$, case A & 0.226 & 0.741 \\
30 $\msun$, case B & 0.217 & 0.750 \\
\enddata
\label{tab:COMassfrac}
\end{deluxetable}
\clearpage

\begin{deluxetable}{lcccccccc}
\footnotesize
\tablecolumns{9}
\tablecaption{
Comparing Stellar Model Parameters Relevant to s-process: 
Maximum Convective Core Mass \& Duration of Helium Burning}
\tablewidth{0pc}
\tablehead{
\colhead{Stellar Model} & \multicolumn{2}{c}{\underline{15 M$_{\sun}$}} &  
                          \multicolumn{2}{c}{\underline{20 M$_{\sun}$}} &
                          \multicolumn{2}{c}{\underline{25 M$_{\sun}$}} &  
                          \multicolumn{2}{c}{\underline{30 M$_{\sun}$}} \\
\colhead{}              & M$_{cc}^{max}$ & $\Delta t_{He}$     &
                          M$_{cc}^{max}$ & $\Delta t_{He}$     &
                          M$_{cc}^{max}$ & $\Delta t_{He}$     &
                          M$_{cc}^{max}$ & $\Delta t_{He}$     \\ 
\colhead{}              & (M$_{\sun}$)   & (10$^6$ yrs)      &
                          (M$_{\sun}$)   & (10$^6$ yrs)      &
                          (M$_{\sun}$)   & (10$^6$ yrs)      &
                          (M$_{\sun}$)   & (10$^6$ yrs)      
}
\startdata
FRANEC (~\protect\cite{K94})  &  2.6  &  1.19  & 4.4  &  0.781 & 
                                 6.5  &  0.620 & 8.6  &  0.525  \\
Schaller et al. (1992)      
                              &  2.80  & 1.12  & 4.83  &  0.79  &
                                 7.13  & 0.63  &   -   &    -      \\
Chieffi et al. (1998)         
                              &   -   &    -   &   -   &    -   &
                                 5.34 &  0.68  &      &   -     \\
Prantzos et al. (1990)
                              &  1.98  &    -   & 3.70  &  1.04  &
                                 5.66  &  0.795 & 7.50  &    -      \\
Present Work 1.7$\times$ $^{12}$C($\alpha$,$\gamma$) CF88            
                              &  2.17  &  1.68  & 3.79  &  1.03  &
                                 5.65  &  0.762 & 7.65  &  0.617    \\
Present Work 1.0$\times$ $^{12}$C($\alpha$,$\gamma$) CF88            
                              &   -   &    -   & 3.71   &  1.01  &
                                  -   &    -   &        &   -     \\
Present Work 1.0$\times$ $^{12}$C($\alpha$,$\gamma$) CFHZ85            
                              &   -   &    -   & 3.85  &  1.08  &
                                  -   &    -   &      &   -     
\enddata
\tablenotetext{}{M$_{cc}^{max}$ is the maximum mass radius of convective core}
\tablenotetext{}{$\Delta t_{He}$  is the duration of core helium burning}
\tablenotetext{}{Parameters of FRANEC results are obtained from figure 10
 of K94 in which $^{12}$C($\alpha$,$\gamma$)$^{16}$O rate of CFHZ85 were used}
\tablenotetext{}{Schaller et al. (1992) use overshooting method during
     the helium convective phase}
\tablenotetext{}{Chieffi et al. (1998) use 
     $^{12}$C($\alpha$,$\gamma$)$^{16}$O rate of CFHZ85}
\tablenotetext{}{Parameters of Prantzos et al. (1990)'s models are obtained
from their Table 1 and Nomoto \& Hashimoto (1988)}  
\label{tab:stellarMod}
\end{deluxetable}

\begin{deluxetable}{lcccccc}
\footnotesize
\tablecolumns{7}
\tablecaption{
   Comparison  of s-Processing in Massive Stars during Core Helium Burning 
   among Authors when $^{22}$Ne($\alpha$,n)$^{25}$Mg rate of
   ~\protect\cite{CF88} is used}
\tablewidth{0pc}
\tablehead{
\colhead{Author} & \colhead{$\tau_c$} &
                   \colhead{n$_{\rm c}^a$} & \colhead{$<\tau>$}  &
                   \colhead{$n_{\rm n}^{max}$} & 
                   \colhead{X$_{22}$}  &
                   \colhead{X$_{80}$/X$_{80 \odot}$$^e$} \\
\colhead{}       &  & 
                   \colhead{}  & \colhead{(mb$^{-1}$)$^b$}  &
                   \colhead{($ \times 10^5$ cm$^{-3}$)$^c$} &
                   \colhead{($ \times 10^{-2}$)$^d$} & \colhead{}
}
\startdata
\multicolumn{7}{c}{\underline{15 M$_{\sun}$} }  \\
P90     & --    & 3.20  &  0.19  &  --    &  1.23  &  72    \\
K94     & --    & 1.85  &  0.09  &  2.05  &  1.65  &  21  \\
A       & 4.00  & 3.38  &  0.10  &  2.27  &  1.33  & 117  \\ 
\multicolumn{7}{c}{} \\
\multicolumn{7}{c}{\underline{20 M$_{\sun}$} } \\ 
P90     & --    & 4.90  &  0.27  &  --    &  0.92  &  342    \\
K94     & --    & 3.66  &  0.15  &  5.06  &  1.32  &  116  \\
A       & 5.93  & 5.48  &  0.16  &  3.50  &  1.04  &  598  \\ 
C       & 6.69  & 6.47  &  0.17  &  2.10  &  0.81  &  1010 \\ 
D       & 6.02  & 5.85  &  0.16  &  3.80  &  0.94  &  719  \\ 
\multicolumn{7}{c}{} \\
\multicolumn{7}{c}{\underline{25 M$_{\sun}$} } \\ 
P90     & --    & 6.20  &  0.33  &  --    &  0.68  &  786    \\
K94     & --    & 5.41  &  0.20  &  6.62  &  1.00  &  475  \\
A       & 7.15  & 6.70  &  0.22  &  4.24  &  0.76  & 1100  \\ 
\multicolumn{7}{c}{} \\
\multicolumn{7}{c}{\underline{30 M$_{\sun}$} } \\ 
P90     & --    & 7.10  &  0.37  &  --    &  0.53  &  1177    \\
K94     & --    & 6.55  &  0.23  &  6.74  &  0.79  &  933  \\
A       & 8.09  & 7.36  &  0.22  &  4.44  &  0.65  & 1368   
\enddata
\tablenotetext{}{P90 is the results from \protect\cite{P90} with
$^{12}$C($\alpha$,$\gamma$)$^{16}$O rate of CF88.}
\tablenotetext{}{K94 is the results from \protect\cite{K94} with
$^{12}$C($\alpha$,$\gamma$)$^{16}$O rate of CFHZ85 (their Table 5).}
\tablenotetext{}{A, C, \& D are the results of the present work.} 
\tablenotetext{}{$\tau_c$ is the central neutron exposure according to
   Eq. (3) in the text.}
\tablenotetext{a}{Number of neutrons captured per iron seed averaged over
the maximum convective core mass.}
\tablenotetext{b}{Mean Neutron exposure at 30 keV, averaged over the
core as $\tau_c$.}
\tablenotetext{c}{Maximum of the mean neutron density.} 
\tablenotetext{d}{Final $^{22}$Ne mass fraction.}
\tablenotetext{e}{Final $^{80}$Kr production factor averaged over the
    maximum convective core mass.}
\label{tab:sproccompareCF88}
\end{deluxetable}

\begin{deluxetable}{lcccccc}
\footnotesize
\tablecolumns{7}
\tablecaption{
   Comparison  of s-Processing in Massive Stars during Core Helium Burning 
   among Authors when the tentative $^{22}$Ne($\alpha$,n)$^{25}$Mg rate of
   ~\protect\cite{K94} is used}
\tablewidth{0pc}
\tablehead{
\colhead{Author} & \colhead{$\tau_c$} &
                   \colhead{n$_{\rm c}^a$} & \colhead{$<\tau>$}  &
                   \colhead{$n_{\rm n}^{max}$} & 
                   \colhead{X$_{22}$}  &
                   \colhead{X$_{80}$/X$_{80 \odot}$$^e$} \\
\colhead{}       &  & 
                   \colhead{}  & \colhead{(mb$^{-1}$)$^b$}  &
                   \colhead{($ \times 10^5$ cm$^{-3}$)$^c$} &
                   \colhead{($ \times 10^{-2}$)$^d$} & \colhead{}
}
\startdata
\multicolumn{7}{c}{\underline{15 M$_{\sun}$} }  \\
K94     & --    & 6.61  &  0.23  &  3.25  &  1.10  &  967  \\
B       & 9.09  & 11.1  &  0.28  &  2.41  &  0.40  &  3883  \\ 
\multicolumn{7}{c}{} \\
\multicolumn{7}{c}{\underline{20 M$_{\sun}$} } \\ 
K94     & --    & 10.7  &  0.31  &  4.93  &  0.50  &  3895  \\
B       & 10.42 & 13.8  &  0.28  &  2.16  &  0.19  &  4918  \\ 
\multicolumn{7}{c}{} \\
\multicolumn{7}{c}{\underline{25 M$_{\sun}$} } \\ 
K94     & --    & 12.9  &  0.35  &  4.29  &  0.17  &  5771  \\
B       & 11.74 & 14.0  &  0.31  &  2.02  &  0.10  &  5332  \\ 
\multicolumn{7}{c}{} \\
\multicolumn{7}{c}{\underline{30 M$_{\sun}$} } \\ 
K94     & --    & 13.6  &  0.36  &  3.74  &  0.07  &  6379  \\
B       & 14.50 & 14.4  &  0.33  &  2.28  &  0.07  &  5498   
\enddata
\tablenotetext{}{All entries have their meanings as in Table 5.} 
\tablenotetext{}{B are the results of the present work with
1.7 $\times$ $^{12}$C($\alpha$,$\gamma$)$^{16}$O rate of CF88.} 
\label{tab:sproccompareK94}
\end{deluxetable}
\clearpage

\begin{deluxetable}{lccrrrrrrrr} 
\footnotesize 
\tablecolumns{11}
\tablecaption{Overproduction factors f=X/X$_{\sun}$ within the maximum
convective core for nuclei in the network
with Z$\leq$40 and f$>$0.05 and some nuclei with 
Z$>$40 and f$>$10}
\tablewidth{0pt} 
\tablehead{ 
\colhead{isotope}  & \colhead{Proton} & \colhead{Mass} & 
\multicolumn{4}{c}{f using $^{22}$Ne($\alpha$,n)$^{25}$Mg of CF88} & 
\multicolumn{4}{c}{f using $^{22}$Ne($\alpha$,n)$^{25}$Mg of WT K94} \\ 
\colhead{}  & \colhead{Number} & \colhead{Number} & 
              \colhead{15 M$_{\sun}$}  & \colhead{20 M$_{\sun}$} &   
              \colhead{25 M$_{\sun}$}  & \colhead{30 M$_{\sun}$} &
              \colhead{15 M$_{\sun}$}  & \colhead{20 M$_{\sun}$} &   
              \colhead{25 M$_{\sun}$}  & \colhead{30 M$_{\sun}$} 
}
\startdata 
c12	&	6	&	12	&	82.3	&	75.3	&	72.4	&	75.4	&	71.6	&	71.4	&	70.0	&	72.1	\\
o16	&	8	&	16	&	74.0	&	75.6	&	76.6	&	76.7	&	76.9	&	77.4	&	77.8	&	77.6	\\
o18	&	8	&	18	&	5.5	&	2.0	&	4.8	&	0.0	&	7.1	&	4.1	&	0.2	&	0.0	\\
ne20	&	10	&	20	&	1.2	&	1.9	&	2.9	&	3.7	&	1.3	&	2.1	&	3.1	&	4.1	\\
ne21	&	10	&	21	&	3.4	&	3.6	&	3.9	&	4.2	&	5.0	&	4.2	&	3.6	&	3.2	\\
ne22	&	10	&	22	&	107.2	&	79.2	&	60.0	&	51.6	&	39.0	&	14.5	&	6.9	&	3.9	\\
na23	&	11	&	23	&	7.6	&	8.0	&	8.1	&	8.2	&	7.7	&	7.4	&	7.3	&	7.3	\\
mg24	&	12	&	24	&	0.9	&	0.9	&	0.9	&	1.0	&	0.8	&	0.8	&	0.9	&	1.0	\\
mg25	&	12	&	25	&	44.4	&	68.6	&	81.5	&	89.0	&	125.5	&	146.7	&	153.5	&	157.3	\\
mg26	&	12	&	26	&	41.9	&	78.1	&	101.0	&	114.3	&	103.9	&	133.7	&	142.6	&	147.3	\\
al27	&	13	&	27	&	0.7	&	0.8	&	0.9	&	0.9	&	0.9	&	1.0	&	1.1	&	1.1	\\
si28	&	14	&	28	&	0.9	&	0.8	&	0.8	&	0.8	&	0.8	&	0.7	&	0.7	&	0.7	\\
si29	&	14	&	29	&	2.1	&	2.3	&	2.3	&	2.4	&	2.1	&	2.2	&	2.2	&	2.2	\\
si30	&	14	&	30	&	1.9	&	2.4	&	2.6	&	2.7	&	2.4	&	2.6	&	2.6	&	2.7	\\
p31	&	15	&	31	&	7.0	&	9.9	&	11.4	&	12.3	&	15.8	&	17.5	&	18.2	&	18.6	\\
s32	&	16	&	32	&	0.6	&	0.5	&	0.5	&	0.4	&	0.4	&	0.4	&	0.4	&	0.3	\\
s33	&	16	&	33	&	30.3	&	31.6	&	31.5	&	31.7	&	27.8	&	26.9	&	26.5	&	26.6	\\
s34	&	16	&	34	&	4.0	&	5.6	&	6.3	&	6.7	&	7.5	&	7.9	&	8.1	&	8.2	\\
s36	&	16	&	36	&	1.1	&	1.2	&	1.9	&	2.9	&	1.6	&	1.7	&	2.0	&	2.2	\\
cl35	&	17	&	35	&	4.6	&	7.5	&	9.0	&	9.8	&	12.2	&	13.6	&	14.0	&	14.3	\\
cl37	&	17	&	37	&	59.2	&	69.5	&	74.6	&	78.1	&	89.8	&	98.5	&	101.4	&	103.4	\\
ar36	&	18	&	36	&	0.3	&	0.2	&	0.2	&	0.1	&	0.1	&	0.1	&	0.1	&	0.1	\\
ar38	&	18	&	38	&	1.3	&	1.6	&	1.8	&	1.9	&	2.3	&	2.6	&	2.6	&	2.7	\\
ar40	&	18	&	40	&	3.2	&	7.2	&	8.5	&	6.9	&	13.2	&	22.9	&	20.4	&	21.5	\\
k39	&	19	&	39	&	1.2	&	1.4	&	1.5	&	1.6	&	2.0	&	2.2	&	2.3	&	2.3	\\
k40	&	19	&	40	&	281.9	&	327.2	&	359.2	&	379.1	&	470.5	&	522.6	&	545.5	&	556.4	\\
k41	&	19	&	41	&	15.2	&	13.3	&	12.9	&	12.8	&	13.7	&	14.5	&	14.9	&	15.1	\\
ca40	&	20	&	40	&	0.4	&	0.2	&	0.2	&	0.2	&	0.1	&	0.1	&	0.1	&	0.1	\\
ca42	&	20	&	42	&	36.5	&	31.8	&	29.1	&	28.2	&	22.7	&	21.4	&	21.0	&	20.9	\\
ca43	&	20	&	43	&	46.7	&	43.8	&	40.9	&	40.1	&	30.8	&	29.1	&	28.6	&	28.5	\\
ca44	&	20	&	44	&	10.4	&	13.9	&	14.6	&	15.1	&	14.5	&	13.9	&	13.8	&	13.8	\\
ca46	&	20	&	46	&	0.7	&	1.8	&	1.7	&	1.1	&	1.8	&	2.7	&	2.4	&	2.7	\\
ca48	&	20	&	48	&	0.9	&	0.8	&	0.8	&	0.8	&	0.7	&	0.7	&	0.7	&	0.7	\\
sc45	&	21	&	45	&	31.6	&	45.4	&	49.1	&	51.4	&	48.8	&	47.4	&	47.0	&	47.2	\\
ti46	&	22	&	46	&	14.6	&	25.2	&	28.9	&	30.9	&	34.2	&	33.5	&	33.3	&	33.5	\\
ti47	&	22	&	47	&	4.6	&	9.0	&	10.8	&	11.8	&	12.9	&	13.0	&	13.0	&	13.1	\\
ti48	&	22	&	48	&	0.6	&	1.4	&	1.8	&	2.0	&	2.5	&	2.5	&	2.6	&	2.6	\\
ti49	&	22	&	49	&	8.0	&	18.9	&	25.7	&	29.3	&	43.4	&	46.5	&	47.7	&	48.5	\\
ti50	&	22	&	50	&	18.0	&	32.7	&	44.9	&	51.7	&	104.9	&	130.4	&	139.1	&	143.2	\\
\tablebreak   
v51	&	23	&	51	&	0.5	&	0.8	&	1.2	&	1.4	&	2.6	&	3.4	&	3.7	&	3.8	\\
cr52	&	24	&	52	&	0.4	&	0.3	&	0.3	&	0.3	&	0.3	&	0.4	&	0.4	&	0.4	\\
cr53	&	24	&	53	&	0.4	&	0.3	&	0.3	&	0.3	&	0.3	&	0.3	&	0.4	&	0.4	\\
cr54	&	24	&	54	&	16.3	&	15.6	&	14.9	&	14.8	&	12.7	&	12.9	&	12.9	&	13.1	\\
mn55	&	25	&	55	&	0.2	&	0.1	&	0.1	&	0.1	&	0.1	&	0.1	&	0.1	&	0.1	\\
fe56	&	26	&	56	&	0.3	&	0.2	&	0.1	&	0.1	&	0.1	&	0.1	&	0.1	&	0.0	\\
fe57	&	26	&	57	&	3.9	&	2.4	&	1.9	&	1.6	&	0.9	&	0.8	&	0.7	&	0.6	\\
fe58	&	26	&	58	&	101.0	&	81.4	&	70.2	&	65.6	&	41.2	&	33.8	&	31.3	&	30.3	\\
co59	&	27	&	59	&	34.6	&	31.2	&	27.9	&	26.6	&	16.7	&	14.0	&	13.0	&	12.6	\\
ni60	&	28	&	60	&	7.5	&	8.0	&	7.6	&	7.4	&	5.0	&	4.3	&	4.0	&	3.9	\\
ni61	&	28	&	61	&	52.2	&	59.4	&	57.4	&	57.0	&	39.1	&	33.4	&	31.4	&	30.9	\\
ni62	&	28	&	62	&	30.8	&	40.6	&	41.4	&	42.2	&	30.8	&	26.8	&	25.5	&	25.2	\\
ni64	&	28	&	64	&	117.0	&	226.0	&	269.9	&	293.4	&	314.6	&	302.5	&	298.5	&	300.4	\\
cu63	&	29	&	63	&	57.5	&	80.5	&	82.8	&	84.7	&	65.8	&	57.3	&	54.1	&	53.6	\\
cu65	&	29	&	65	&	99.5	&	193.1	&	231.2	&	251.1	&	265.8	&	252.5	&	249.3	&	251.0	\\
zn64	&	30	&	64	&	14.2	&	22.4	&	24.2	&	25.4	&	20.4	&	18.3	&	17.6	&	17.6	\\
zn66	&	30	&	66	&	44.8	&	101.9	&	130.4	&	145.6	&	171.4	&	169.2	&	169.5	&	171.4	\\
zn67	&	30	&	67	&	62.1	&	147.6	&	192.1	&	216.3	&	261.5	&	260.5	&	261.8	&	265.0	\\
zn68	&	30	&	68	&	54.2	&	158.4	&	225.1	&	262.3	&	402.4	&	429.0	&	440.8	&	449.0	\\
zn70	&	30	&	70	&	3.7	&	13.5	&	21.6	&	26.0	&	58.9	&	70.3	&	74.1	&	75.8	\\
ga69	&	31	&	69	&	63.5	&	194.5	&	281.4	&	330.0	&	525.2	&	566.1	&	583.8	&	595.2	\\
ga71	&	31	&	71	&	76.3	&	268.0	&	409.3	&	489.2	&	868.2	&	971.1	&	1012.3	&	1034.8	\\
ge70	&	32	&	70	&	81.6	&	269.6	&	401.5	&	475.4	&	811.8	&	891.3	&	924.2	&	943.4	\\
ge72	&	32	&	72	&	54.2	&	210.1	&	334.3	&	404.0	&	804.0	&	924.8	&	971.6	&	995.3	\\
ge73	&	32	&	73	&	33.0	&	133.0	&	215.5	&	262.8	&	519.9	&	609.2	&	643.7	&	660.6	\\
ge74	&	32	&	74	&	26.7	&	117.7	&	197.9	&	242.9	&	554.8	&	666.3	&	707.8	&	727.0	\\
ge76	&	32	&	76	&	0.1	&	0.1	&	0.1	&	0.0	&	0.2	&	0.2	&	0.1	&	0.1	\\
as75	&	33	&	75	&	19.4	&	87.3	&	148.3	&	182.8	&	419.7	&	508.6	&	542.0	&	557.4	\\
se76	&	34	&	76	&	55.8	&	259.7	&	447.5	&	552.5	&	1342.2	&	1640.1	&	1750.0	&	1799.7	\\
se77	&	34	&	77	&	23.0	&	109.0	&	189.5	&	234.8	&	576.5	&	709.8	&	759.3	&	781.6	\\
se78	&	34	&	78	&	28.0	&	141.0	&	252.6	&	314.9	&	840.0	&	1059.3	&	1139.0	&	1173.9	\\
se80	&	34	&	80	&	2.3	&	13.5	&	24.7	&	31.2	&	95.2	&	129.4	&	135.4	&	139.4	\\
br79	&	35	&	79	&	8.1	&	41.1	&	74.1	&	92.4	&	248.1	&	314.0	&	337.0	&	347.0	\\
br81	&	35	&	81	&	3.7	&	28.0	&	45.0	&	36.5	&	154.3	&	221.3	&	210.0	&	194.2	\\
kr80	&	36	&	80	&	117.3	&	606.3	&	1100.5	&	1374.6	&	3883.1	&	4946.6	&	5337.9	&	5509.6	\\
kr82	&	36	&	82	&	54.9	&	304.5	&	578.4	&	731.0	&	2329.6	&	3099.9	&	3368.3	&	3485.6	\\
kr83	&	36	&	83	&	15.4	&	88.1	&	171.1	&	217.5	&	697.7	&	950.2	&	1039.2	&	1076.6	\\
kr84	&	36	&	84	&	12.1	&	71.8	&	147.8	&	189.4	&	722.1	&	1044.9	&	1158.2	&	1204.2	\\
kr86	&	36	&	86	&	1.0	&	5.7	&	8.0	&	6.6	&	49.3	&	143.3	&	116.5	&	132.0	\\
rb85	&	37	&	85	&	8.1	&	48.3	&	100.2	&	127.3	&	508.6	&	741.6	&	815.2	&	845.8	\\
rb87	&	37	&	87	&	0.7	&	2.5	&	3.7	&	2.7	&	24.2	&	71.8	&	65.3	&	74.9	\\
\tablebreak 
sr86	&	38	&	86	&	46.8	&	273.9	&	587.6	&	764.1	&	3369.9	&	5121.8	&	5758.8	&	6016.6	\\
sr87	&	38	&	87	&	40.2	&	230.6	&	507.9	&	662.0	&	3152.6	&	4932.2	&	5587.7	&	5841.2	\\
sr88	&	38	&	88	&	13.2	&	57.5	&	127.7	&	167.6	&	1085.8	&	2007.4	&	2311.3	&	2438.9	\\
y89	&	39	&	89	&	9.8	&	35.5	&	76.0	&	99.5	&	756.2	&	1550.1	&	1794.5	&	1903.2	\\
zr90	&	40	&	90	&	4.3	&	13.6	&	27.3	&	35.1	&	298.6	&	673.8	&	781.8	&	833.1	\\
zr91	&	40	&	91	&	5.1	&	15.5	&	30.4	&	39.0	&	332.7	&	793.9	&	924.5	&	987.8	\\
zr92	&	40	&	92	&	4.4	&	12.4	&	22.9	&	28.8	&	252.4	&	641.5	&	746.1	&	800.0	\\
zr94	&	40	&	94	&	3.4	&	8.6	&	14.7	&	18.1	&	151.4	&	422.7	&	490.2	&	528.9	\\
zr96	&	40	&	96	&	0.2	&	0.1	&	0.1	&	0.1	&	0.1	&	0.5	&	0.3	&	0.3	\\
mo96	&	42	&	96	&	3.1	&	7.7	&	12.8	&	15.6	&	124.9	&	361.8	&	420.2	&	454.7	\\
mo98	&	42	&	98	&	2.2	&	5.2	&	8.5	&	10.2	&	78.8	&	233.7	&	270.5	&	293.2	\\
ru100	&	44	&	100	&	2.7	&	6.0	&	9.7	&	11.7	&	87.6	&	263.8	&	305.0	&	331.0	\\
pd104	&	46	&	104	&	2.6	&	5.6	&	8.9	&	10.6	&	74.9	&	230.5	&	266.2	&	289.6	\\
gd152	&	64	&	152	&	22.4	&	35.3	&	43.1	&	47.3	&	64.8	&	75.5	&	79.6	&	81.9	\\
ta180	&	73	&	180	&	 1.0	&	 4.5	&	  9.4	& 7.4	&	 3.3	&	 11.1	&	 16.6	&	 15.3	
\enddata
\label{tab:overprod}
\end{deluxetable}

%% file: figures.tex
\clearpage
\setcounter{figure}{0}
\begin{figure}
\plotfiddle{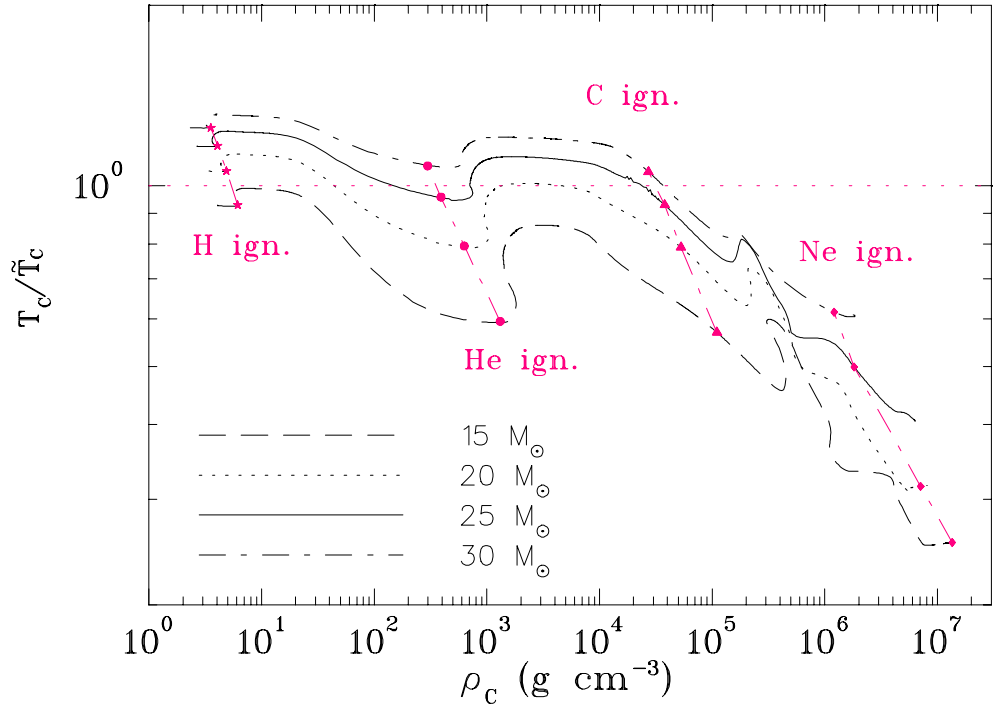}{240pt}{90}{70}{70}{240}{-80}
\vspace{0.5in}
\caption{
Evolution of the central temperature vs. central density for the four case
A stellar models.  
The temperature scale, 
$\tilde{T}_C$ is $\approx$ 2$\times$10$^7 \; \rho^{1/3}$.
All four star models have been evolved up to central neon ignition.
\label{fig:TCvsRhoC}
}
\end{figure}
\clearpage

\begin{figure}
\makebox[5.00in]{\psfig{figure=f2.eps,angle=90,width=5.0in}}
\caption{
The convective helium-burning core mass in the four case A star models versus
the fraction of the helium-burning time (upper panel) and the mass fraction
of $^4$He (lower panel).
Starting at X($^4$He) = 0.1, the edges of the convective cores have been
finely resolved to prevent a sudden injection of a large supply of
$^4$He into the cores.
\label{fig:MconvXHe}
}
\end{figure}
\clearpage

\begin{figure}
\makebox[5.60in]{\psfig{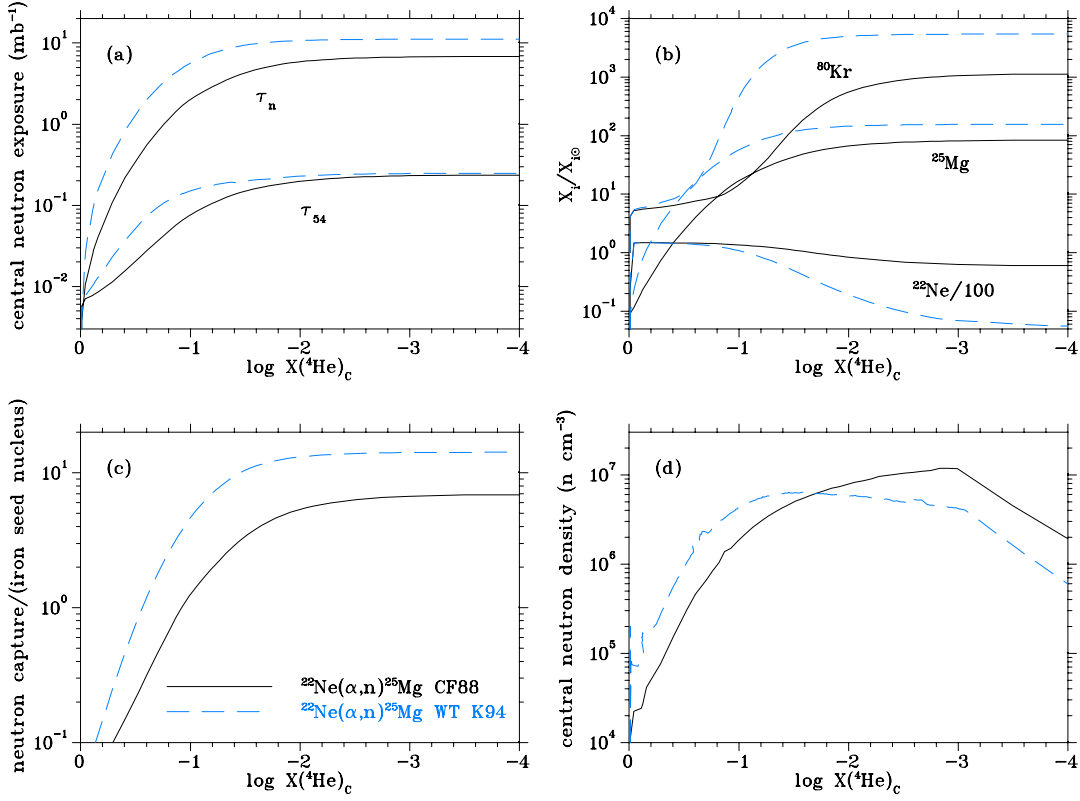}}
\caption{
Several s-process diagnostics for two 25 $M_\odot$ stellar models.
The case A model is shown as the solid curve while the case B model
is shown as the dashed line.
The diagnostics are the following: 
a) $\tau_n$ is the central neutron exposure, as computed from Eq.
(\protect\ref{eq:taun}).  It is the neutron exposure that would have been
seen by a nucleus that stayed
in the center of the star throughout the burning ($\tau_n$).
$\tau_{54}$  is the neutron
exposure computed from the decline of the $^{54}$Fe abundance
due to neutron captures (see Eq. [\protect\ref{eq:tau54}]).  
The latter better indicates the average neutron exposure
of the core helium-burning s-process because it reflects the dilution by
the convective
mixing.  It also shows that the bulk of the s-processing occurs mainly in
the range X($^{4}$He)$_C$ $\simeq$ 0.01 - 0.001.
b) Overabundances of key isotopes. Note that the production factor of
$^{22}$Ne is multiplied by 1/100 in order that it not overlap 
in the figure that of $^{25}$Mg.
c) The number of neutrons captured per iron-seed nucleus according to Eq.
(\protect\ref{eq:nc}). 
d) The central neutron density of the star during core helium burning.  
All quantities are plotted against central helium mass fraction.
A low-lying $^{22}$Ne($\alpha$,n)$^{25}$Mg resonance (case B) would 
dramatically increase the robustness of the core helium-burning s-process.
\label{fig:centralsproc25Msun}
}
\end{figure}
\clearpage

\begin{figure}[t]
\makebox[5.0in]{\psfig{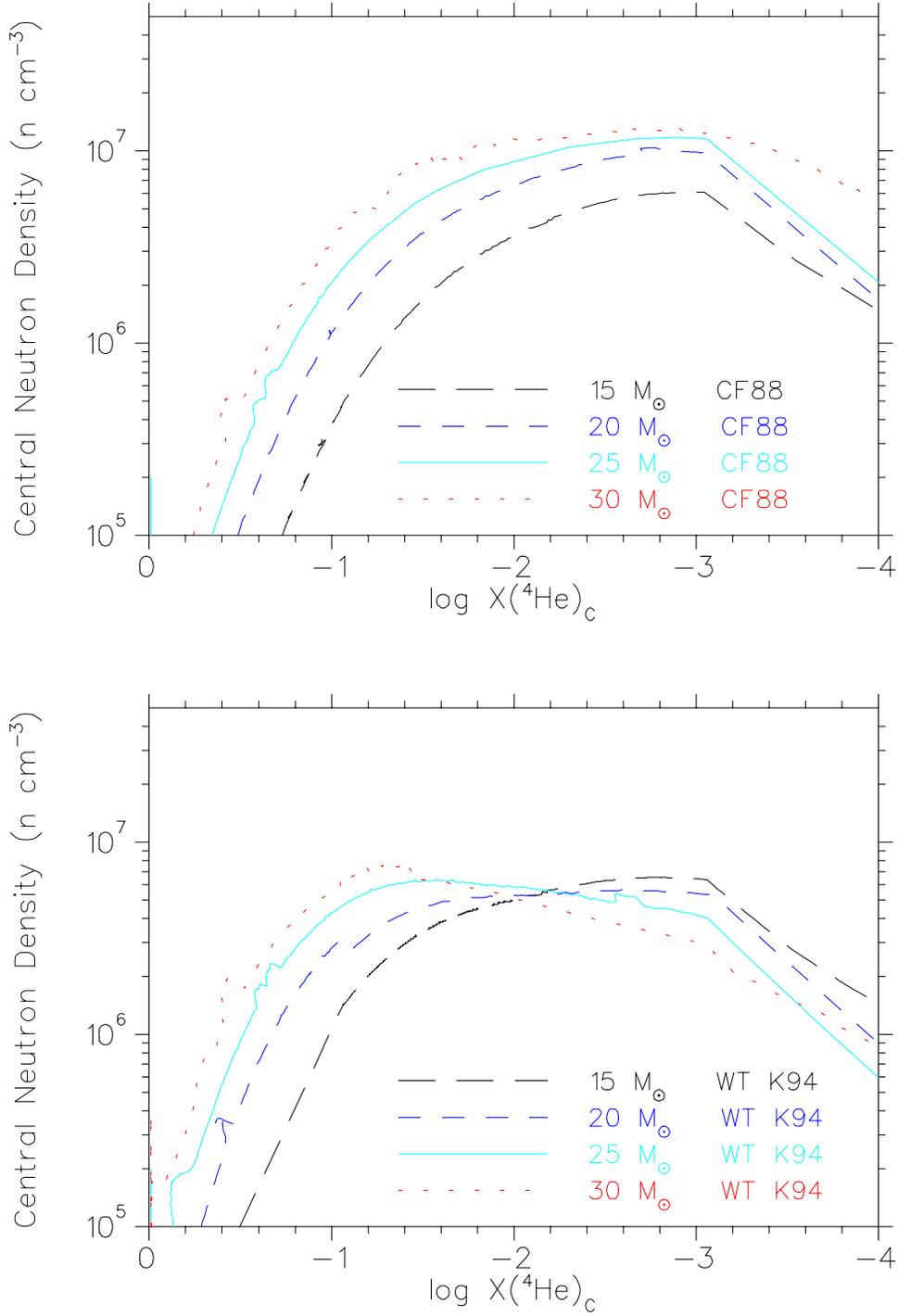}}
\caption{
The central neutron density as function of central helium mass fraction
or time for the four case A (top panel) and the four case B (lower panel)
stellar models.
\label{fig:centndnsvsXHe}
}
\end{figure}
\clearpage

\begin{figure}[t]
\makebox[6in]{\psfig{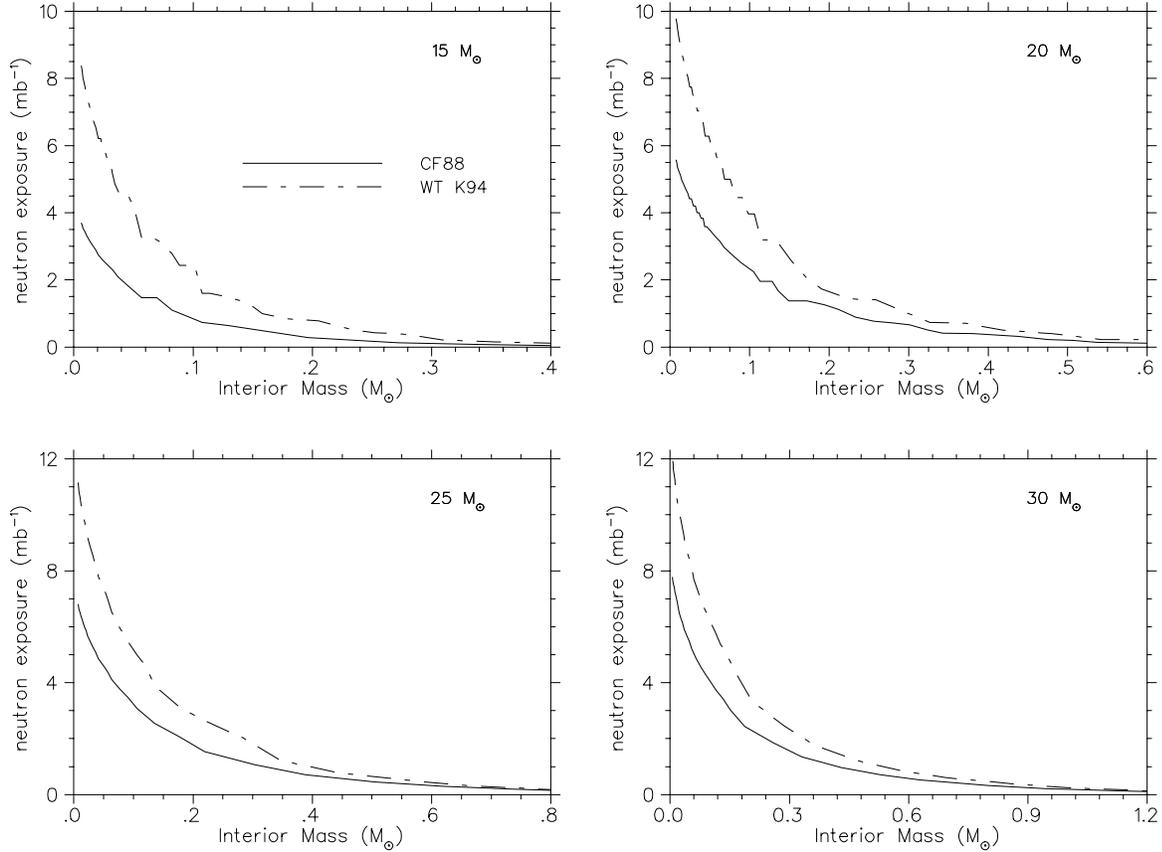}}
\caption{
The neutron exposure $\tau_n$ as a function of interior mass for all
eight case A (solid curve) and B (dashed curve)
stellar models through core helium burning.  
$\tau_n$ is the neutron exposure a nucleus would experience were it
to remain at the particular interior mass.
This figures clearly demonstrates that most neutron captures occur
in the very center of the star.  
Nuclei are convected down into the center of the star, capture neutrons, 
and then mix back out into the core.
\label{fig:tauneutvsMr}
}
\end{figure}
\clearpage

\begin{figure}
\makebox[6.00in]{\psfig{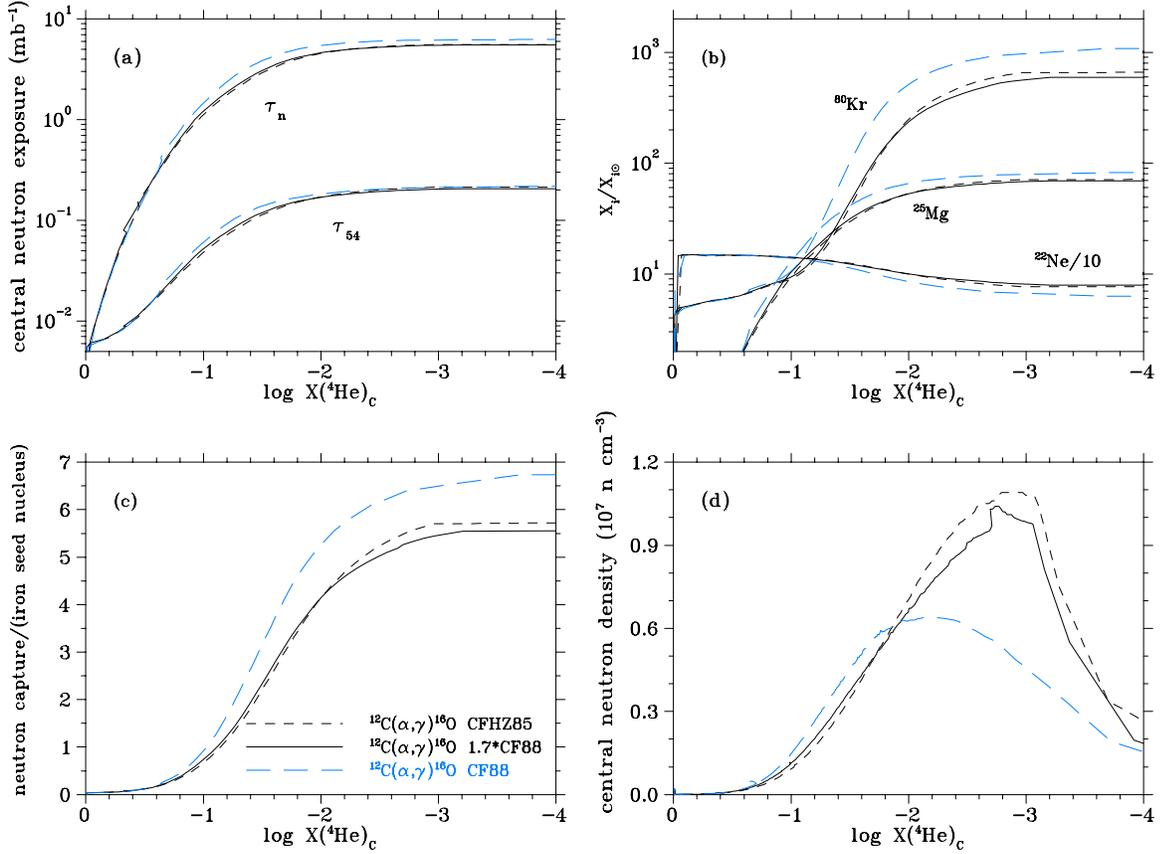}}
\caption{Similar to Figure \protect\ref{fig:centralsproc25Msun} but for
the case A (solid curve), case C (long-dashed curve), and case D (short-dashed
curve) 20 M$_{\sun}$ stellar models.
The lowest $^{12}{\rm C}(\alpha,\gamma)^{16}$O reaction rate gives the
largest degree of s-processing.  The results of cases A and D agree to
within $\sim 10-20\%$.
\label{fig:centralsproc20Msun}
}
\end{figure}
\clearpage

\begin{figure}
\makebox[6.00in]{\psfig{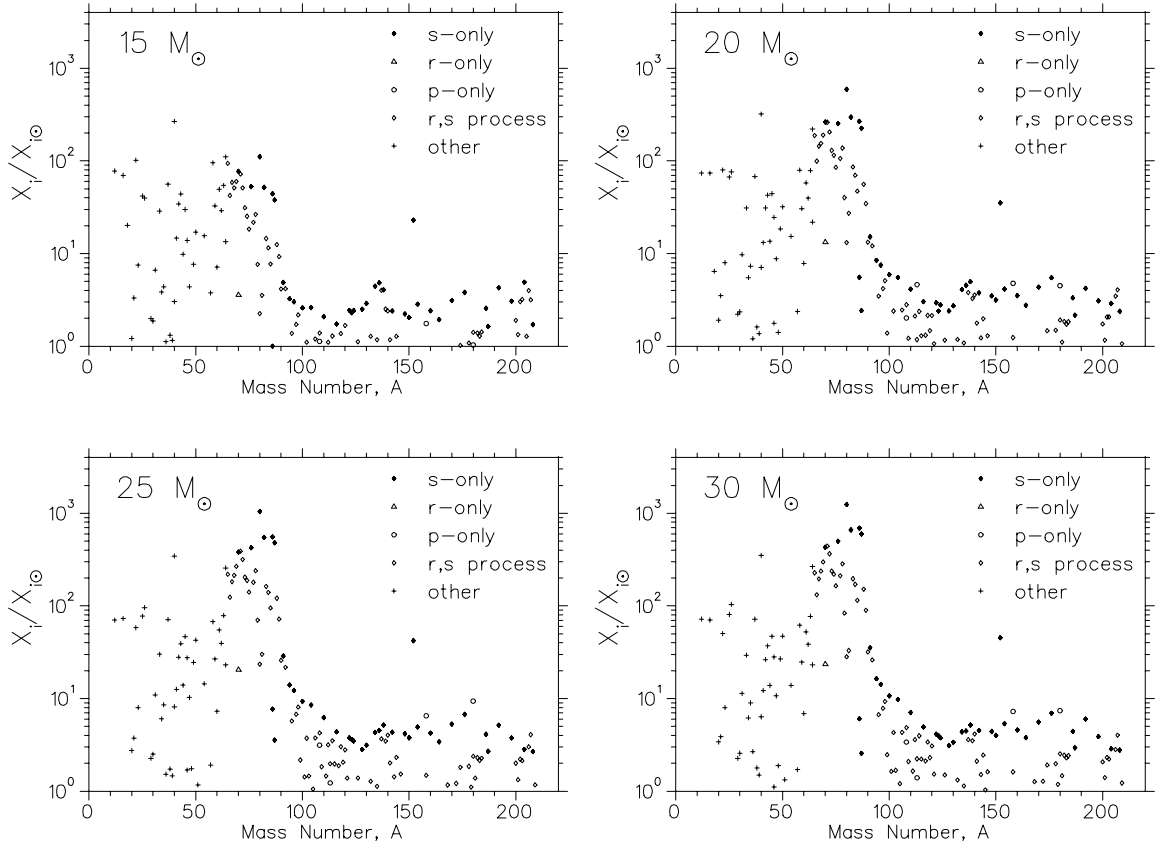}}
\caption{
Overabundances of heavy nuclei (mass number $A\geq 12$) averaged over the
convection helium-burning core for the four case A stellar models.
The primary nucleosynthesis production process for each
isotope is indicated by the symbol type.
\label{fig:prodfactRCF88}
}
\end{figure}
\clearpage

\begin{figure}
\makebox[6.00in]{\psfig{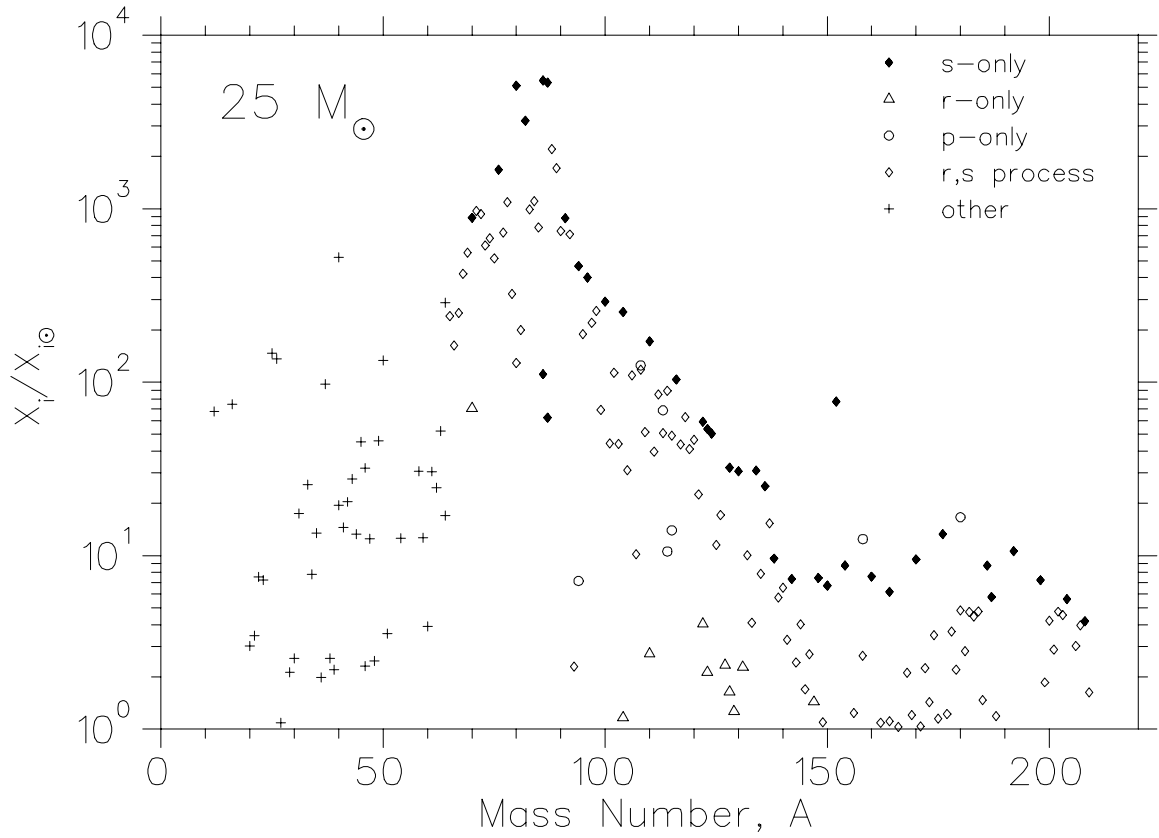}}
\caption{
Overabundances of heavy nuclei (mass number $A\geq 12$) averaged over the
convection helium-burning core for the 25 $M_\odot$ stellar model with
10\% contribution to the reaction rate for $^{22}{\rm Ne}(\alpha,n)^{25}$Mg
from a low-lying resonance.
\label{fig:prodfact25}
}
\end{figure}
\clearpage

\begin{figure}[t]
\makebox[2.8in]{\psfig{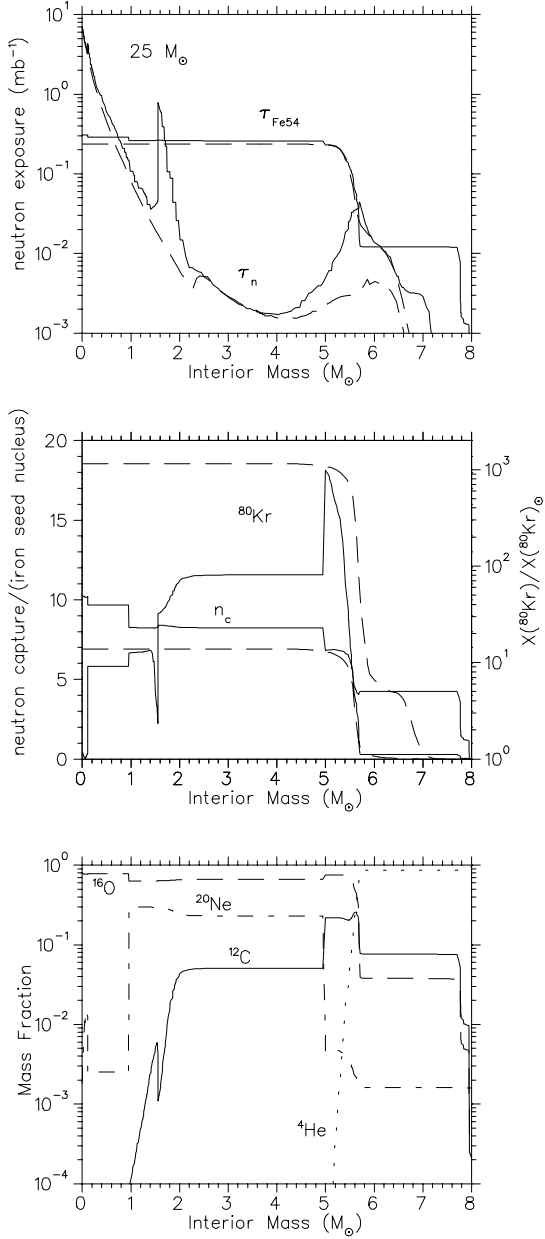}}
\caption{
Several s-process diagnostics for the 25 $M_\odot$ case A stellar model
at central neon ignition.  
The upper panel shows the two measures of the neutron exposure.
Of particular interest is the run of $\tau_n$ against interior mass.
The dashed line shows the conditions at the end of core helium burning,
while the solid line shows the conditions at the end of core neon burning.
Most s-processing has occurred in the center of the star, but the contribution
from shell carbon burning and shell helium burning are evident at 1.6
$M_\odot$ and about 5.4 $M_\odot$, respectively.  
The middle panel shows the
number of neutrons captured per iron seed nucleus and the overabundance of
$^{80}$Kr at the end of core helium (dashed line) and neon burning (solid
line).  
Various convective burning phases after central helium burning
modify the number of neutrons captured per seed nucleus.
Interestingly, shell carbon s-processing has destroyed a significant
fraction of the $^{80}$Kr produced in earlier convection core helium
burning.
The bottom panel shows mass fractions at the end of core neon burning
of some nuclei to indicate the nuclear burning region.
\label{fig:kr80Cshell}
}
\end{figure}
\clearpage

\clearpage